\documentclass[fleqn,usenatbib]{mnras}

\usepackage{newtxtext,newtxmath}

\usepackage[T1]{fontenc}

\DeclareRobustCommand{\VAN}[3]{#2}
\let\VANthebibliography\thebibliography
\def\thebibliography{\DeclareRobustCommand{\VAN}[3]{##3}\VANthebibliography}

\usepackage{graphicx}	
\usepackage{amsmath}	

\title[Unveiling the structural content of NGC 6357]{Unveiling the structural content of NGC 6357 via kinematics and NIR variability}

\author[C. Ordenes-Huanca et al.]{
C. Ordenes-Huanca,$^{1, 2, 3}$\thanks{E-mail: ccordenes@uc.cl}
M. Zoccali,$^{1, 2}$
A. Bayo,$^{3, 4, 5}$
J. Cuadra,$^{3, 6}$
R. Contreras Ramos$^{1,2}$
\newauthor
and A. Rojas-Arriagada$^{7,2,8,9}$
\\
$^{1}$Instituto de Astrofísica, Pontificia Universidad Católica de Chile, Casilla 306, Santiago 22, Chile\\
$^{2}$Millenium Institute of Astrophysics (MAS), Nuncio Monseñor Sótero Sanz 100, Providencia, Santiago Chile\\
$^{3}$N\'ucleo Milenio Formaci\'on Planetaria - NPF, Chile\\
$^{4}$European Southern Observatory, Karl-Schwarzschild-Strasse 2, 85748 Garching bei M\"{u}nchen, Germany\\
$^{5}$Instituto de F\'isica y Astronom\'ia, Facultad de Ciencias, Universidad de Valpara\'iso, Av. Gran Breta\~na 1111, Valpara\'iso, Chile\\
$^{6}$Departamento de Ciencias, Facultad de Artes Liberales, Universidad Adolfo Ib\'a\~nez, Av. Padre Hurtado 750, Vi\~na del Mar, Chile \\
$^{7}$Departamento de F\'isica, Universidad de Santiago de Chile, Av. V\'ictor Jara 3659, Santiago, Chile\\
$^{8}$N\'ucleo Milenio ERIS\\
$^{9}$Center for Interdisciplinary Research in Astrophysics and Space Exploration (CIRAS), Universidad de Santiago de Chile, Santiago, Chile\\
}

\date{Accepted XXX. Received YYY; in original form ZZZ}

\pubyear{2023}

\begin{document}
\label{firstpage}
\pagerange{\pageref{firstpage}--\pageref{lastpage}}
\maketitle

\begin{abstract}
NGC 6357, a star-forming complex at $\sim 1.7$ kpc from the Sun, contains giant molecular clouds and three prominent star clusters alongside with HII regions, very massive stars and thousands of young stellar objects in different evolutionary stages. We present a combined infrared kinematic and time domain study of the line of sight towards this region enabled by the VVVX survey. In terms of kinematics, a novel discovery emerges: an asymmetrical distribution in the vector point diagram. Some stars in the sample exhibit spatial proximity to dusty regions, with their proper motions aligned with filament projections, hinting at a younger population linked to triggered star formation. However, this distribution could also stem from an asymmetric stellar expansion event within NGC 6357, warranting further investigation. Comparing this data with Gaia revealed inconsistencies likely due to high extinction levels in the region. Additionally, owing to accretion episodes and surface cool spots, young stars display high variability. Using the $K_s$-band time series data, we overcome the extreme levels of extinction towards the region, and compile a catalogue of $774$ infrared light curves of young stars. Each light curve has been characterized in terms of asymmetry and periodicity, to infer the dominant underlying physical mechanism. These findings are then correlated with evolutionary stages, aiming to uncover potential age disparities among the observed stars. This study contributes to our understanding the intricate dynamics and evolutionary processes within NGC 6357, offering valuable insights into the formation and development of stellar populations within such complex environments.

\end{abstract}




\section{Introduction}
\label{sec:intro}

The study of different star-forming regions and their stellar populations, as well as the reconstruction of their star-formation history and structure need astrometric and kinematic data with the lowest levels of uncertainty possible. Thanks to the \textit{Gaia} satellite \citep{Gaia_2016, Gaia_2018, Gaia_2023}, information about the position, parallaxes, proper motions (PMs) and radial velocities (RVs), in addition to photometric data, of more than a billion stars have become available in the last few years. The movements of these objects in a six-dimensional phase space allow one to find co-moving groups that belong to the same cluster and helps to identify actual members of each of those structures. Likewise, these data indicate the origin and dynamical evolution of young star clusters \citep{Kuhn_2019}. The latter is very important because each stellar population could retain the kinematic signature from the places in which it was born \citep{Wright_2020}.\\

Moreover, star-forming regions that are close to the Sun, such as Upper Scorpius and Taurus, have been kinematically studied in order to expand the number of members of each one or to identify clustered substructures (see e. g. \citealt{Roccatagliata_2020, BricenoMorales_2023}). These are very suitable sites to study using \textit{Gaia} data, because, in addition to their short distance from the Sun, they also possess low levels of extinction. Co-moving groups are identified and sub-structures of diverse origin are also recovered.\\

\subsection{From membership to dynamical evolution}

Several efforts to reconstruct the structure and star-formation history of different star-forming regions through kinematic and astrometric data have also been made. For example, \citet{Kounkel2018} presented an extended study of the Orion Complex and its young stellar object (YSO) population, taking advantage of the available \textit{Gaia} DR2 and APOGEE-2 data \citep{Blanton_2017}. The authors reassemble the properties of the parental molecular clouds that formed the different substructures in this region using the kinematic information available. Furthermore, they discuss the dynamical evolution of several clusters that compose this region. Particularly, they found that the region is composed of five kinematically different clusters. In addition, \citet{Zari2019} also utilized \textit{Gaia} DR2 data to study the kinematics, ages and 3D structure of the Orion OB association. Stars with a range of ages and also contrasting kinematic stellar sub-groups are found, allowing the authors to conclude that they could have distinct origin. These highlight the capabilities and improvements that \textit{Gaia} has brought to the understanding of the structures observed towards active star-forming sites in our Galaxy and their birth. However, since Gaia observes in optical wavelengths, these studies are biased towards more evolved objects and/or those located in less extincted areas. Therefore, the kinematics of the youngest, more embedded sources remain unstudied.\\

The study of the Orion Complex could be performed using optical data because it is a fairly nearby star-forming region, located at approximately $\sim 400$ pc, that includes sites devoid or with modest levels of extinction \citep{Hillenbrand1998, Uehara2021}. The study of more distant star-forming regions is important because much of the stellar production originates at distances farther away than Orion. So that, limiting our studies to the latter is not representative of the entire process \citep{Evans_2022}.\\ 

Our knowledge of the structure of highly extincted star-forming regions remains scarce, because they are not accessible through visual bands, or at least not their low-mass, less luminous members. This is aggravated when these sites are located at higher distances. Therefore, only a moderate number of stars can be kinematically studied in these regions \citep{Russeil_2020}. Most stars in our Galaxy form in those environments, which motivates the study of sites with these features \citep{Kennicutt_2012}.\\

The Vista Variables of the Via L\'actea survey \citep[VVV and its extension VVVX; hereafter simply VVVX]{minniti+2010} has mapped our Galaxy in the near-IR for approximately eight years. Thus, giving access to very extinct regions of the bulge and part of the disk of the Milky Way. These include several star-forming regions which are usually affected by the presence of dust. The VVVX has a wavelength range appropriate to study the stellar content of more extincted sites that are actively forming stars. In addition, in the $K_s$-band, about $80$ epochs are available, spread across a time baseline of eight years, allowing to account for flux variability, besides PMs. The latter are very relevant data, because they allow us to study less massive young stars, which are much more numerous \citep{OrdenesHuanca2022}. Particularly, it helped us to probe intermediate to low-mass stars in the massive and highly extincted region called NGC 6357.\\

\subsection{NGC 6357 star-forming complex}

NGC 6357 is an active star-forming region located in the Sagittarius arm of the Milky Way, towards its bulge, at a distance of approximately $1.7$ kpc from the Sun, as reported by recent studies \citep{Fang2012, Russeil2012, Lima2014}. It hosts very massive stars \citep{Massey2001} and thousands of YSOs in different evolutionary stages \citep{Wang2007, Broos2013}, which are mainly distributed across three open clusters. One of them is Pismis 24. With an estimated age of $\sim 1$ Myr, it is known to contain two of the brightest stars in our Galaxy, namely Pis 24-1, a multiple system with at least three components, and Pis 24-17 \citep{Bohigas2004}, all of them with masses of $\approx 100$ $M_{\odot}$ \citep{MaizApellaniz_2007}. The other two clusters, AH03J1725–34.4 and the ``B cluster", named by \citet{Massi2015}, were found to contain several sources with IR-excess, which indicates how young the members of this region are, and were found to have approximately the same size ($\sim 2.5$ pc) and number of stars as Pismis 24. For the latter, it was estimated to have $\sim 3600 - 11000$ members. Further, these three clusters are believed to be coeval \citep{Getman_2014}.\\

\begin{figure}
    \includegraphics[width=\columnwidth]{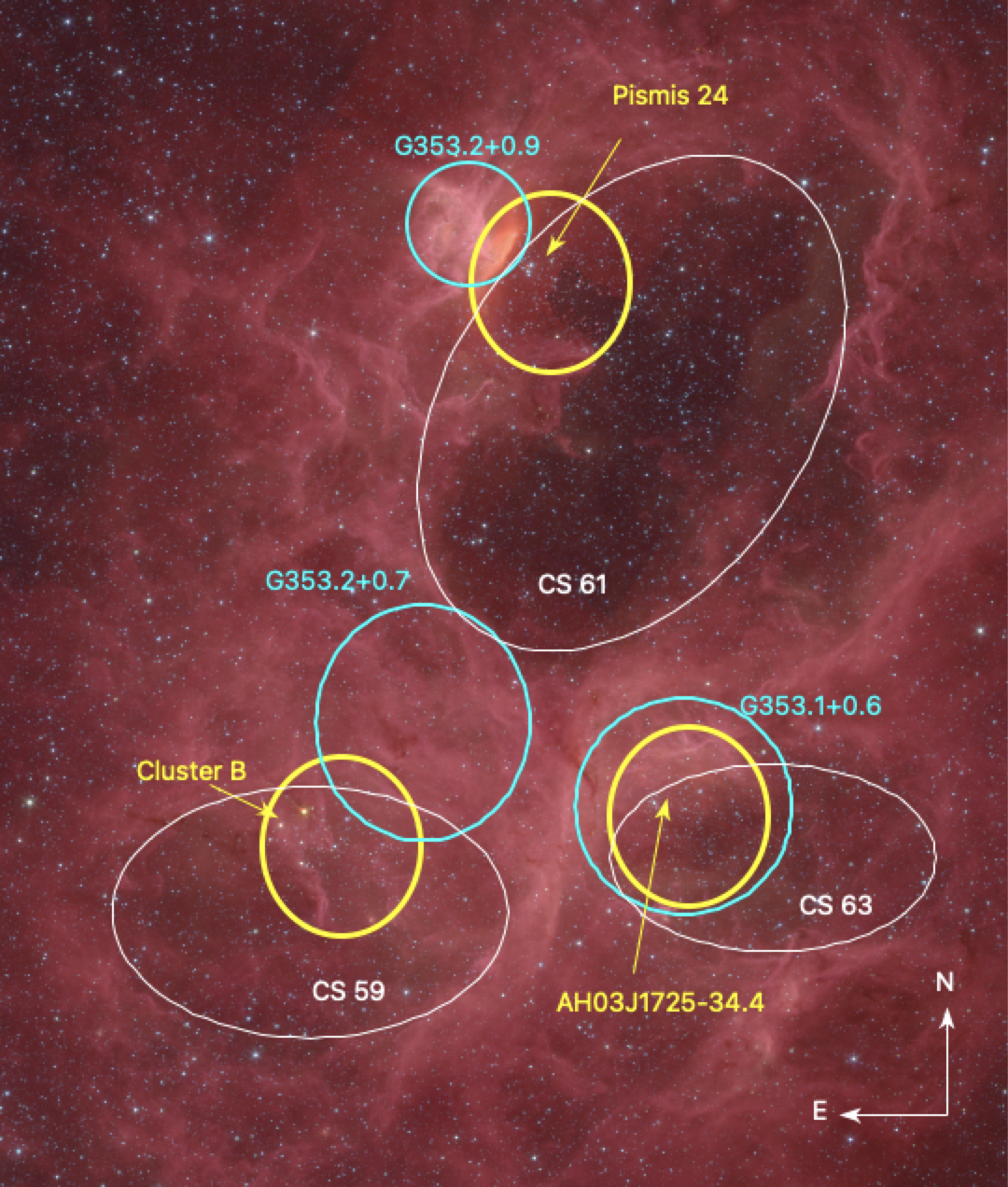}
    \caption{Spitzer RGB image of NGC 6357 ($8$ $\mu m$ in red, $4.2$ $\mu m$ in green and $3.6$ $\mu m$ in blue). The approximate location of several structures are indicated, such as HII regions in cyan circles and bubbles in white ellipses. The position of the three stellar clusters is also marked (yellow ellipses).}
    \label{fig:NGC6357_spitzer}
\end{figure}

Other structures are part of the complex, as shown in the annotated false color image of Spitzer in Fig~\ref{fig:NGC6357_spitzer}. Molecular clouds have also been found in the area as well as, at least, three HII regions. Particularly, G353.2 + 0.7, G353.1 + 0.6 and the youngest and brightest one being G353.2 + 0.9 \citep{Cappa2011, Massi2015}. Each of them are associated with, and excited by, the stellar members of the three open clusters mentioned above. In addition, three bubbles coexist in the area, usually designated as CS 59, CS 61 and CS 63 \citep{Churchwell2007}. These are structures dominated by dust continuum emission, but where there is little or no PAHs emission. Each of them is linked to one of the open clusters of the region. CS 61 and CS 63 are thought to be formed due to winds created by the strong UV field produced by the luminous OB stars in the area. The origin of CS 59 instead, is not yet clear \citep{Fang2012}.\\

\begin{figure}
    \includegraphics[width=\columnwidth]{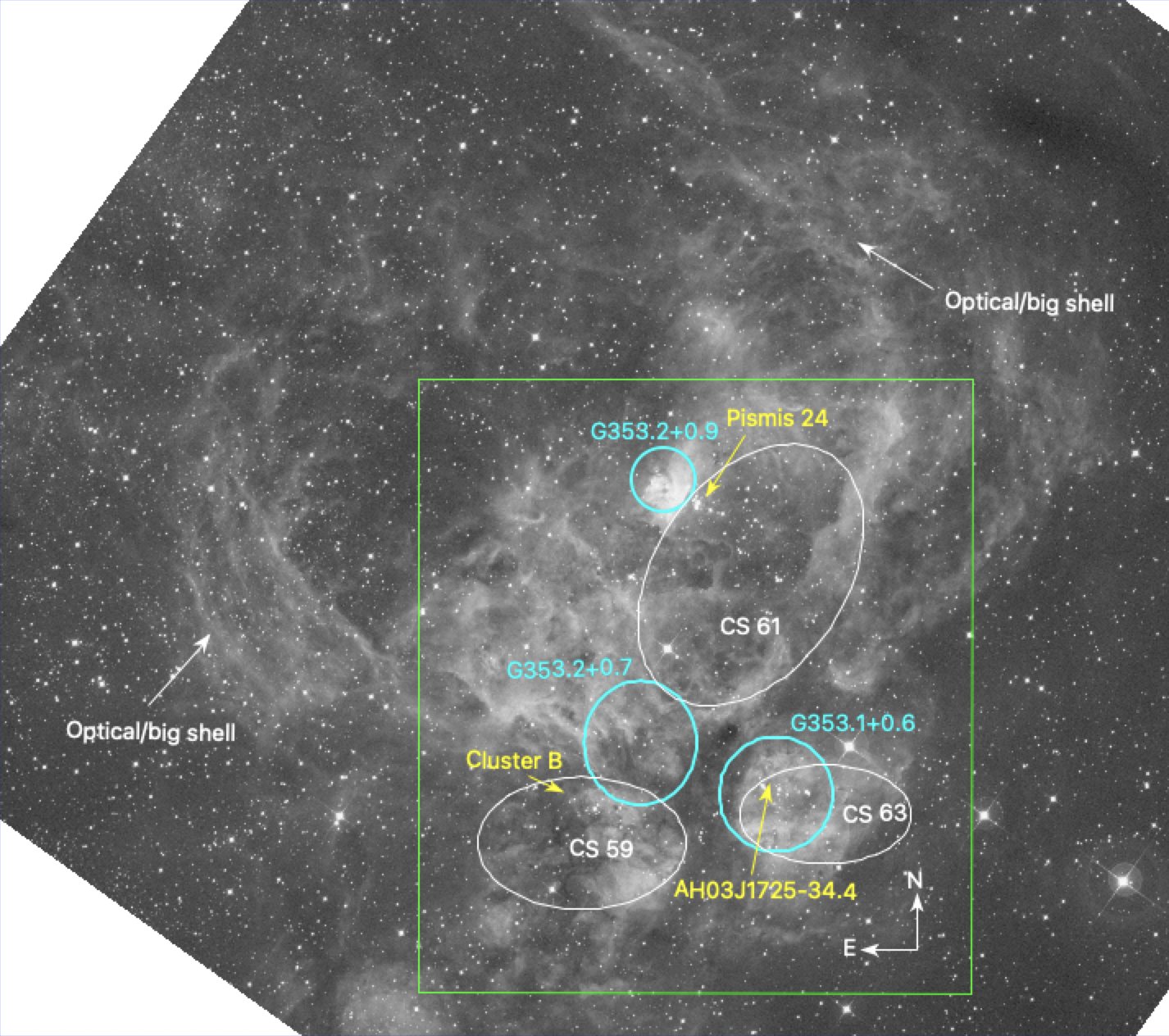}
    \caption{DSS R-band image of NGC 6357. The approximate location of several structures are indicated, as in Fig.~\ref{fig:NGC6357_spitzer}. The green box indicates the region of the Spitzer image and the big shell is also labeled.}
    \label{fig:NGC6357_DSSR}
\end{figure}

In the optical, a ``big shell" or ``ring" has been observed \citep{Cappa2011, Massi2015}. It is opened to the north and has a projected diameter of $\sim 60'$, as shown in Fig.~\ref{fig:NGC6357_DSSR}. Due to its shape, its origin has been attributed to a supernova event \citep{Wang2007}, but this has not yet been confirmed.\\

What is observed in the interstellar medium (ISM) of this region has been interpreted as evidence for triggered star formation. This idea comes from the fact that pillars and shocks have been found in the complex \citep{Westmoquette2010}. These shocks are a consequence of ionization fronts originated by massive stars and induce the collapse of gas cores and clumps \citep{Russeil2010, Giannetti2012}, hence, the formation of new stars \citep{Bohigas2004}. In addition, a member star, namely [PFR86] G353.19+0.91 IRS 4, appears at the tip of a pillar (or ``elephant trunk") and this can also be interpreted as an indicator for triggered star formation. All these facts lead to the inference that a younger population of stars is expected in NGC 6357. However, these younger members have not been confirmed by Hertzsprung-Russell diagrams (HRDs) or by marked differences in the fraction of disk-bearing stars among different structures in the ISM \citep{Fang2012, Massi2015}, which suggests that the age difference between the two generations could be small.\\

In the work by \citet{Russeil_2017}, the authors considered a population of OB stars in the area to investigate age differences between them. These ages were derived considering photometric and spectroscopic data to locate the stars in the HRD and, then, compare their positions with stellar evolutionary models. In their sample, $2$ stars from Pismis 24 had an age of $\sim 1.4$ Myr, whereas $13$ objects from the NGC 6357 region were $\sim 4.6$ Myr old. This suggested that a first star formation event took place at this time, but the main burst occurred $1.4$ Myr ago. The latter is in agreement with the ages found for the low-mass pre-Main Sequence population of stars of the three main clusters of NGC 6357 by \citet{Getman_2014}. In this work, the authors derived the median age of distinct clusters comprising different star-forming regions (and not individual stellar ages). Their age computation was based on a combination between the X-ray luminosity ($L_X$) and NIR magnitudes of the cluster members and strongly relies on a relationship between $L_X$ and the mass of pre-Main Sequence stars. For the clusters of NGC 6357, they found that all these stars have ages in the range of $1 - 1.5$ Myr, Still, these age determinations should be taken carefully. The uncertainties of the ages found by \citet{Russeil_2017} are mentioned as $38 \%$ for O-type stars and $94 \%$ for B stars. On the other hand, \citet{Getman_2014} mentioned that for a typical cluster (of $N \sim 20$ stars), the uncertainty will be $\sim 0.5$ Myr. Nevertheless, all these features make NGC 6357 a very rich complex of gas and dust structures interacting with its stellar content.\\

Unfortunately, the region presents high levels of extinction, with $A_{v}= 5.93 \pm 0.49$ mag \citep{Russeil2012}, complicating the study of its stellar population. Nevertheless, efforts have been made to compile catalogues of the young stars in this area based on different wavelength indicators, such as X-ray emission and IR excess. These members have shown to be predominantly massive Main Sequence stars or intermediate to low-mass stars in the pre-Main sequence phase, in addition to the existence of protostars \citep{Wang2007, Feigelson2013, Povich2013}. The kinematics of the region have also been studied, through the available \textit{Gaia} optical data. \citet{Russeil_2020} investigated the kinematics of OB stars and YSOs members of NGC 6357, along with NGC 6334 and GM1-24. However, the extinction of the region clearly limited the sample of YSOs considered (only $66$ between the three sites).\\

\subsection{Variability of YSOs}

Depending on their mass, pre-Main Sequence stars are known to vary their brightness according to one or two different mechanisms. Due to dynamo effect, magnetic fields in these sources are so strong that they present large dark or cool spots \citep{Petrov2003} in well defined regions of the stellar surface \citep{Grankin_2008, OrdenesHuanca2022}. Spots coupled with rotation lead to periodic brightness variations linked to their rotation rate \citep{Rydgren_1983}. In these cases, periods are of the order of days \citep{carpenter2001, Carpenter_2002}.\\

The second mechanism that applies in principle to all young stars, regardless the mass of the central object, has to do with the presence of a surrounding disk and subsequent accretion events. On one hand, the star accretes material from the inner disk through the magnetic field lines. This, and other processes can create inhomogenities in the dust density of the inner disk that will leave their signatures when crossing our line of sight, leading to decreases in the brightness we observe \citep{Morales_Calder_n_2011, Bodman_2017}.\\

On the other hand, the material that arrives on the star originates bright spots in the stellar surface. Accretion is a stochastic phenomenon by nature and gives rise to mostly irregular flux changes. However, in some cases, bright spots persist for long enough to produce periodic brightness variations when the star rotates \citep{Kesseli_2016}, just like cool spots. Furthermore, it has been demonstrated that a low-mass companion can induce periodic or pulsed accretion events \citep{Dunhill_2015, Teyssandier_2020, Guo_2022, nogueira2023}. Embedded YSOs can also display accretion related flux changes. These have been mainly attributed to eruptive behaviours over a range of amplitudes and at different timescales. Particularly, the source can show long duration outbursts (called FU Orionis or FUors), when accretion rates are enhanced for years \citep{Fischer_2023}. Conversely, the star can present short duration outbursts associated to higher accretion rates maintained from weeks to months (EX Lupi events or EXors) \citep{ContrerasPena2017}. Nevertheless, accretion related variability can also be observed as stochastic and in much shorter timescales \citep{Stauffer_2016}. While this behavior seems to be more frequent on embedded sources, it can also be observed in Classical T Tauri disks \citep{Robinson_2022}.\\

All the physical processes mentioned above will leave a signature on what we see in the light curves of pre-Main sequence stars, which can be observed at different wavelengths, including the near-IR. In this frequency, brightness changes related to the stellar photosphere, as well as the inner disk structure can be detected \citep{Rebull2014, OrdenesHuanca2022}.\\

In this work, we have compiled a catalog of $774$ light curves for intermediate to low-mass young stars that have been already identified as members of the NGC 6357 region. Massive stars are not part of this study because they usually appear saturated on VVVX images, unless they are very extinct. However, $3$ O-type star candidates are present from the work by \citet{Wang2007}. We present, for the first time, brightness variations for these stars over a time span of eight years, the VVVX baseline of observations. Each of them was classified according to their degree of periodicity and asymmetry, following \citet{Cody_2014}. Additionally, for a part of these stars, we could constrain rotation periods. \\

In order to confirm that these sources are true members of this region, we analyzed their PMs and found that, according to the VVVX data, it seems that there are possibly two kinematically different populations. One of them is spatially related to filamentary zones with movements along these structures. If these two populations are actually coexisting, this could imply that a triggered star formation process took place within the evolution of NGC 6357. Nevertheless, this kinematical behavior can also stem for an asymmetric expansion experienced by the stars of the region.\\

This paper is organized as follows: in Sec.~\ref{sec:catalogue}, we present the previous studies included here to produce our inital census of NGC 6357 with a counterpart in VVVX data. The asymmetry observed in the movements of these stars in the plane of the sky and their link to molecular clouds are part of Sec.~\ref{sec:pms}, whereas Sec.~\ref{sec:QM} is devoted to the period search for a subset of members with light curves available. Also, their classification according to asymmetry and periodicity metrics is included. In Sec.~\ref{sec:discussion} we relate the two kinematically different populations to different properties, such as our own light curve classification and parameters from the literature. Finally, in Sec,~\ref{sec:conclusions} we conclude and summarize our results.

\section{NGC 6357 stars found in VVVX data}
\label{sec:catalogue}

\subsection{Stellar members of NGC 6357 from the literature}

The stellar content of NGC 6357 has been studied for several years. Very massive, OB stars have been identified and confirmed as members \citep{Neckel1978} and have been used to find the distance and extinction to this region \citep{Lortet1984}. Other authors have studied stars in a wider range of masses.\\

\citet{Wang2007} analyzed the intermediate to low-mass population of pre-Main sequence stars in Pismis 24 (roughly in the range $\sim 0.3 - 16$ $M_{\odot}$), as well as O-type star candidates, and their environment, using \textit{Chandra} observations. X-rays are known to trace the magnetic activity of stars, which is quite strong in pre-Main Sequence stars compared to Main Sequence (MS) sources and it can vary in time \citep{Damiani_1995}. In addition, \textit{Chandra} observations can reach regions with high levels of extinction, even up to $A_{V}\sim 500$ \citep{Grosso_2005} and are not affected by nebular emission from the HII regions, as optical and IR wavelengths are \citep{Kuhn2021}. Therefore, this allowed the authors to identify the young population of the cluster. However, X-rays are more effective to detect the low-mass members. In the work by \citet{Wang2007}, $779$ stars are presented as part of the open cluster Pismis 24, of which $665$ are labeled as highly reliable sources. The remaining are listed separately as spurious background detections. However, most of them have optical or IR counterparts. Still, contamination by AGNs or field stars can be present in their census. The authors estimate this to be the case for less than 4\% of the sources\\

Despite the fact that X-rays are a strong indicator of magnetic activity, which is a recurrent feature in a young stellar population, the identification of members only by this parameter could be biased to Class III objects with only low-mass disks, if any. Pre-Main Sequence stars that are still accreting material have proven to be less luminous in X-rays \citep{Prisinzano2008}. As young stars can also host disks or envelopes, this is translated to the observation of an infrared excess in their SEDs. Moreover, very massive stars, which are usually located inside a cluster \citep{Broos2013}, can be identified in the optical, where their low-mass and faint counterparts are difficult to detect. All these facts lead to expect that a multiwavelength approach to confirm young members is more appropriate.\\

The Massive Young star-forming Complex Study in Infrared and X-rays project (MYStIX) characterized $20$ near ($d \leq 4$ kpc) young clusters, dominated by the presence of OB stars, and their environments \citep{Feigelson2013}. This work combined optical, IR and X-ray data, from \textit{Chandra X-ray Observatory}, \textit{Spitzer Space Telescope} and the United Kingdom InfraRed Telescope (UKIRT) complemented by 2MASS data and archival optical catalogues of OB stars, to present and identify members of massive star-forming regions, including NGC 6357. In their approach, X-ray emission and variability was considered and combined with other parameters, such as, the $J$-band magnitude and the mid- and near-IR excess, to select the young stellar population of a given star-forming region. The result was the identification of hundreds of OB stars and $31784$ low-mass pre-Main Sequence stars spread over the $20$ star-forming regions considered \citep{Broos2013}. Particularly, for NGC 6357, $2235$ stellar members were identified.\\ 

As previously stated, a mostly X-ray driven census can be biased against heavily accreting objects (either from their envelopes or disks). However, another factor which can affect the detection of young objects at this frequency is that this is usually a variable radiation. There could be a population of these objects that is in a low state of X-ray emission and that will be left out of the MYStIX catalogue. Therefore, the identification of an IR excess in a given source is an important parameter to classify such star as a YSO.\\ 

\citet{Povich2013} presented the MYStIX InfraRed Excess Source (MIRES) catalogue, as a complement of MYStIX. Its goal was to look for mid- and near-IR excess emission from sources in the same star-forming regions considered by MYStIX. Here, only IR data from Spitzer/IRAC, 2MASS and UKIRT were studied, taking a larger field of view than MYStIX. Using these, the SEDs of the objects observed were constructed and analyzed. The ones that resembled those from a young star were pointed as YSO candidates. The MIRES catalogue presented sources with IR excess from $1 - 8$ $\mu m$, increasing the possibility of finding objects in a low X-ray emission state, which could have been left off the MYStIX list. For NGC 6357, $545$ stars were flagged as members and classified as likely YSO in the MIRES catalogue.\\

For a far star-forming region, MIRES contains data mainly of intermediate-mass YSOs with masses between $2 - 8$ $M_{\odot}$, but it is incomplete at solar masses. Furthermore, if there are a large number of field stars towards a given star-forming region, contamination from giant stars with dusty envelopes may be present.\\

Considering the three catalogues mentioned above, we constructed a base literature catalogue of NGC 6357 members. Particularly, for the MIRES one, we only included sources flagged as members and likely YSOs. In addition, from MYStIX, we only included sources classified as young stars in the massive star-forming region. By combining the three catalogues and removing potentially duplicated sources that are closer together than $1"$, we ended up with $2534$ unique young members of NGC 6357.\\

\subsection{Cross-match between the literature catalogue and VVVX data}

The literature catalogue (hereafter \textit{LitCat}) of $2534$ stars was cross-matched to our VVVX data, using a $1.5"$ matching radius. The $1.5"$ figure was chosen as a compromise between efficiency of source recovery and avoidance of false cross-identification. Even though the astrometric precision of VVV is about $25$ mas for stars of $K_{s} = 15$ mag \citep{Saito_2012}, the same level of accuracy is not available for the literature coordinates. In addition, the VVVX survey is seeing limited, so we gauge the average seeing ($\sim 1"$) in our radius optimization. Taking this into account, we found $2243$ counterparts out of which only $1883$ had light curves available in the $K_s$-band. These compose the literature census detected in VVVX data (hereafter \textit{LitCatVVVX}).\\

A fraction of sources from the literature ($277$) were matched with more than one star from VVVX data. In these cases, we considered the brightest star in the $K_s$ data to be its most likely counterpart.\\


Conversely, there were stars in the VVVX data that were matched to more than one star from the literature ($82$). In these cases, we decided to keep the star from the literature that were closest to the VVVX one and consider it as its counterpart. It is important to mention that, in the great majority of cases, the closest star was also the brightest. Only $2$ stars did not meet this condition. So that, by considering the closest star, we are also keeping the brightest star as its VVVX counterpart. In addition, all selected stars shown signs of
flux variability, a behavior fully expected from young stars.\\

Since blending can induce artificial variability, we visually inspected the epochs of the LitCatVVVX objects with the lowest seeing values. Sources that appeared with close companions, extended or very close to a bright star were removed from the list, leaving us with $1786$ stars. As a note of caution, a part of these stars can appear blended on higher seeing images, which could affect their variability.\\

In order to analyze the kinematic properties of members from the literature, we had to apply a further filter requiring that the detections had sufficient quality for the VVVX pipeline to estimate their PMs \citep{contrerasramos17}. To do so, we considered $1634$ out of $1786$ stars that had available PMs on the VVVX data set. These comprise the final sample for the kinematic analysis. All the filter mentioned are summarized in Table~\ref{tab:conditions}. In addition, all the stars with available PM values in VVVX data, along with their galactic coordinates and photometric data are listed on Table \ref{tab:pms}.\\

\begin{table}
	\centering
	\caption{Conditions applied to the base literature catalogue of young stars with the number of objects that met the condition.}
	\label{tab:conditions}
	\begin{tabular}{cc} 
		\hline
		Condition & \# stars\\
		\hline
            LitCat & 2534\\
		LitCatVVVX & 1883\\
		No close or bright companions, point-like & 1786\\
            Stars with VVVX PMs & 1634\\
		\hline
	\end{tabular}
\end{table}

\section{Kinematics of NGC 6357}
\label{sec:pms} 

\subsection{PMs from VVVX data}

For stars with $K_{s}<15$ mag, VVVX PMs have a mean precision of $\sim 0.51$ mas $yr^{-1}$ in each Galactic coordinate. Their computation is explained in \citet{contrerasramos17}.\\

The NGC 6357 sample contains PM values with mean uncertainties $\overline{e\mu_{l}\cos b} \approx 0.55$ mas $yr^{-1}$ and $\overline{e\mu_{b}} \approx 0.53$ mas $yr^{-1}$, in the Galactic longitude and latitude direction, respectively. In addition, only $9 \%$ of them have $K_{s}>15$ mag.\\

\begin{figure}
    \centering
    \includegraphics[width=\columnwidth]{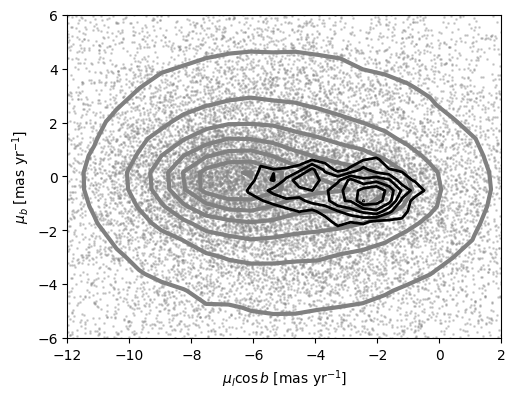}
    \caption{VPD for NGC 6357 stars with VVVX counterpart that have available PM values (black density contours). Gray points represent stars in the same region of NGC 6357 with density contours superimposed.}
    \label{fig:pms}
\end{figure}

The sample of $1634$ stars with PM values in VVVX data includes $7$ O-type star candidates from \citet{Wang2007} and Pismis 24-12, a B-type star of $11 M_{\odot}$ \citep{Fang2012}. Their location in the vector point diagram (VPD) is shown as black contours in Fig.~\ref{fig:pms}. Stars with available PMs in VVVX within a rectangular FoV of $352.5 \leq l \leq 353.6$ and $0.3 \leq b \leq 1.2$ are shown as gray points with density contours added. \\

YSOs in the VVVX data seem to show a higher dispersion in $\mu_{l}\cos b$ when compared to $\mu_{b}$. These stars clearly deviate from the motion of the majority of stars in the same line of sight. \\

\begin{figure}
    \centering
    \includegraphics[width=0.95\columnwidth]{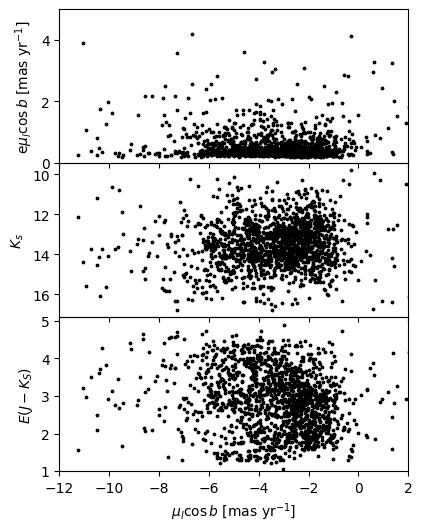}
    \caption{\textit{Upper panel}: Error in the Galactic coordinate PM e$\mu_{l} \cos b$ vs. $\mu_{l} \cos b$. \textit{Middle panel}: $K_s$ magnitude vs. $\mu_{l} \cos b$. \textit{Bottom panel}: Color excess $E(J-K_s)$ vs. $\mu_{l} \cos b$.}
    \label{fig:mul_dependance}
\end{figure}

In order to check if this difference in the projected movement on the sky has an artificial origin, we tried to correlate the $\mu_{l} \cos b$ values with their uncertainties, the average magnitude in the $K_s$-band of the sources and the color $E(J-K_s)$ affecting them (pair plots shown in Fig.~\ref{fig:mul_dependance}). The latter is obtained from the data of \citet{Surot_2020}. No obvious correlations were found among any of these comparisons. The Spearman correlation coefficient was computed for the three dependencies, obtaining values near zero. Fig.~\ref{fig:mul_dependance} shows that the stars with large negative $\mu_{l}\cos b$ do not have larger errors (top panels), nor fainter $K_s$ magnitudes (middle panels). For the color $E(J-K_s)$, however, it seems that a near anti-correlation could be present, particularly for stars with less negative $\mu_{l} \cos b$ values, which look to be associated with lower reddening. Its Spearman correlation coefficient has a higher value compared to the other two dependencies ($\approx 0.20$). This is further discussed later in the text.\\

\subsection{Independent diagnostic: comparison with Gaia PMs}

As mentioned in Sec.~\ref{sec:intro}, \textit{Gaia} data is available towards the line of sight of NGC 6357. Using Gaia DR2, \citet{Russeil_2020} made a dynamical study of this site and NGC 6334. Here, a very limited sample of YSOs and OB stars was considered to investigate past dynamical history.\\

\begin{figure}
    \includegraphics[width=0.9\columnwidth]{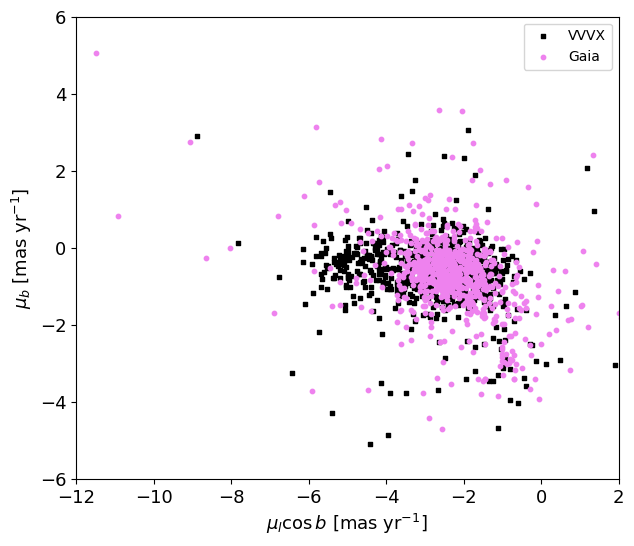}
    \caption{VPD for the $731$ common objects between Gaia (violet open circles) and VVVX (black squares).}
    \label{fig:vpd_gaia_vvvx}
\end{figure}

We made a similar study using Gaia DR3 data \citep{Gaia_2023}, finding $731$ common objects with PMs measured in both datasets. The comparison between VVVX and Gaia PMs is in the VPD of Fig.~\ref{fig:vpd_gaia_vvvx}. In this plot, Gaia equatorial PMs were converted to the Galactic reference frame. Here, a clear difference is observed for the most negative $\mu_{l} \cos b$ values. Particularly, the $\mu_{l} \cos b$ dispersion observed in the VVVX VPD is not present on the Gaia PMs. As in \citet{Russeil_2020}, only one overdensity is observed in the optical. However, it seems that a other small overdensity appears in their data (at pmRA $\approx -1$ mas $yr^{-1}$ and pmDEC $\approx -3$ mas $yr^{-1}$.) The mean Gaia PMs found for all the stars is very similar to the ones presented in \citet{Russeil_2020}, with $\langle \mu_{l} \cos b \rangle _{Gaia} \approx -2.36$ mas $yr^{-1}$ and $\langle \mu_{b} \rangle_{Gaia} \approx -0.79$ mas $yr^{-1}$. We attribute this discrepancy to the high levels of extinction of NGC 6357, which may prevent to have reliable values of PMs for all the sources in the area using optical data. A similar discrepancy is also found in \citet{AlonsoGarcia_2021} for 2MASS-GC 02, the most reddened of the clusters they analyzed.\\

\begin{figure}
    \includegraphics[width=0.9\columnwidth]{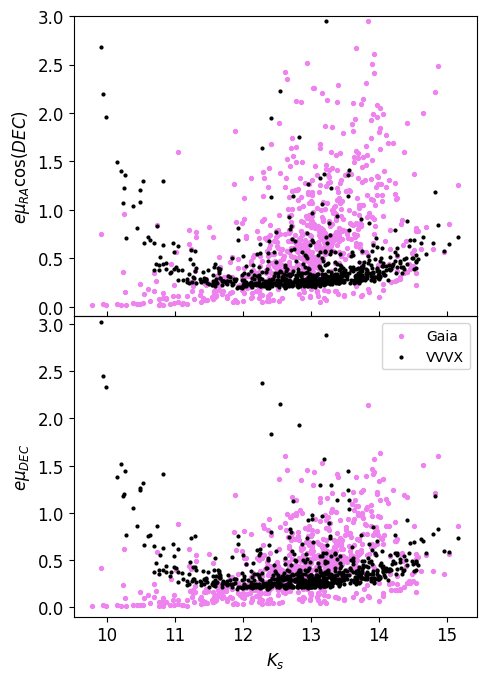}
    \caption{Errors in PM measurements, in each equatorial coordinate (RA in upper panel and DEC in bottom panel) against the $K_s$ magnitude of the common stars between Gaia (violet dots) and VVVX (black dots).}
    \label{fig:epms_comparison}
\end{figure}

In Fig.~\ref{fig:epms_comparison}, the relation between errors in the PM measurements of the $731$ stars in common with Gaia against the $K_s$ magnitude is shown. In these plots, we are considering equatorial coordinates PMs to match with Gaia. One can observe that the error values in Gaia PMs (violet dots) are higher on average than the ones in VVVX data (black dots). This is particularly true for stars with magnitudes between $12 \leq K_s \leq 14$, which make up the majority of our data set. Gaia PM errors have lower values only for brightest stars for which VVVX is saturated. In addition, Gaia PMs are known to be affected by systematics, so they could also be underestimated \citep{Lindegren_2021}. As errors can also be related to the dispersion of the points in the VPD, we computed the standard deviation of each dataset in the Galactic latitude coordinate $\sigma(\mu_{b})$, as the Galactic longitude one should be more affected by the presence of the possible two kinematic overdensities. We obtained that, for VVVX values, $\sigma (\mu_{b})$ is $\approx 1.14$ mas $yr^{-1}$, whereas for Gaia is $\sigma(\mu_{b}) \approx 1.38$ mas $yr^{-1}$, demonstrating the wider dispersion of the latter.\\




In \citet{Luna_2023}, the authors also mention that Gaia PM errors (and their RUWE values) could be underestimated in crowded regions towards the Galactic bulge, such as globular clusters. The surface density of the clusters that belong to NGC 6357 has been estimated as $800$ stars pc$^{-2}$ \citep{Fang2012}. This is an order of magnitude lower than that of globular clusters, but the PM errors of Gaia can still be underestimated at this surface density value, according to the results of \citet{Luna_2023}.\\

\begin{figure}
    \centering
    \includegraphics[width=0.9\columnwidth]{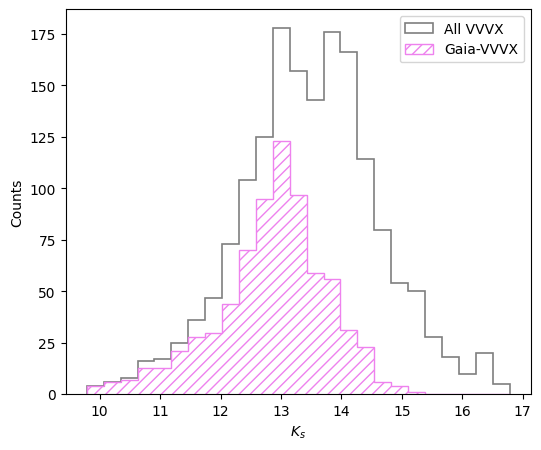}
    \caption{$K_s$ magnitude distribution of stars with VVVX PM measurements (gray) and the one for the $731$ stars in common between VVVX and Gaia (violet).}
    \label{fig:ks_comparison}
\end{figure}

In addition, the distribution of the $K_s$ magnitudes of all stars for which VVVX PM values are available is presented in Fig.~\ref{fig:ks_comparison} as the gray histogram. Of those stars, only the ones at the bright side are in common with Gaia (violet hatched histogram). Fainter and redder objects are not recovered in the optical. This and the PM error dispersion could cause the observed difference between the projected movement of stars in the near IR and optical bands. Further, the majority of the $731$ stars in common between VVVX and Gaia have magnitudes $G>18$, so they are faint in the optical. VVVX data is crucial to observe both populations of stars and properly describe this star-forming region.\\

In order to better understand the source of the discrepancy between the PMs from VVVX and those from Gaia, we checked for possible systematic errors affecting VVVX PMs. First, we verfieid that the two overdensities in the VPD are observed within each chip, excluding problems in the astrometric calibration of different chips. Second, we verified that stars with offset PMs do not coincide with those with envelopes or disks, according to their $Q_{JHHK_s}$ reddening-free index. Those stars might have larger FWHM in $K_s$ resulting in an offset centroid with respect to the optical. We found that these objects are not necessarily showing the largest differences between their PMs in the optical compared to the ones in the NIR. This index and the disk/evelope candidates are presented in the next subsection.\\

\begin{figure}
    \centering
    \includegraphics[width=0.8\columnwidth]{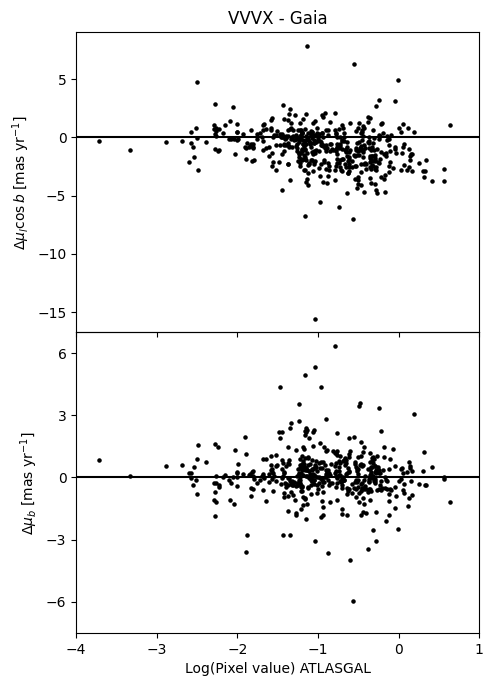}
    \caption{\textit{Top panel:} Difference in the PM values for the Galactic longitude coordinate, $\Delta \mu_{l} \cos b$, against the pixel value from ATLASGAL. \textit{Bottom panel:} Same as top panel, but for the PM difference in the Galactic latitude coordinate, $\Delta \mu_{b}$. The horizontal black line is located at $\Delta \mu = 0$ in each Galactic coordinate.}
    \label{fig:mu_atlas_comparison}
\end{figure}

We then tried to investigate whether Gaia's PMs could have a bias restricted to the stars with large extinction. To this end, we tried to correlate the difference in PM values from VVVX and Gaia with the logarithm of the pixel value obtained from ATLASGAL, in $870$ $\mu$m, where higher values indicate higher dust continuum emission. Here, the position of a given star was associated to the nearest pixel from the ATLASGAL image. These results are shown in Fig.~\ref{fig:mu_atlas_comparison} and confirm that for higher values of dust emission, the PM difference is higher. This is particularly true for $\Delta \mu_{l} \cos b$, where a higher difference is observed at approximately Log(Pixel value)$= 0$ and beyond (top panel). The Spearman correlation coefficient between these two parameters is $\approx -0.31$, indicating a moderate correlation. The horizontal black line is located at $\Delta \mu = 0$ in each Galactic coordinate.\\

\begin{figure}
    \centering
    \includegraphics[width=0.8\columnwidth]{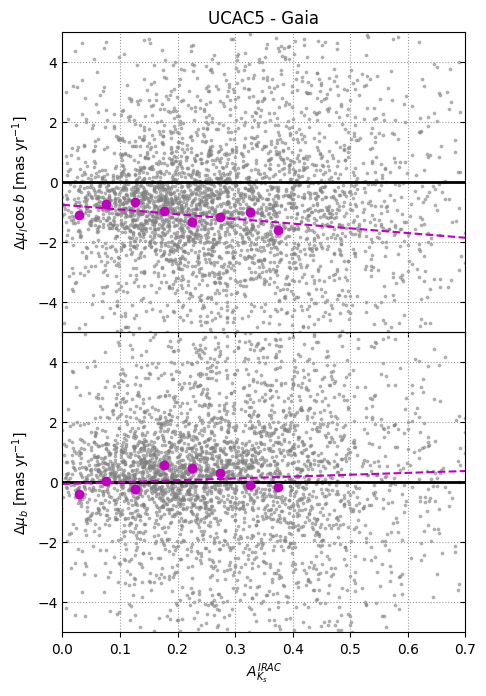}
\caption{\textit{Top panel:} Difference in the PM values for the Galactic longitude coordinate, $\Delta \mu_{l} \cos b$, against the extinction $A_{K_s}^{IRAC}$. \textit{Bottom panel:} Same as top panel, but for the PM difference in the Galactic latitude coordinate, $\Delta \mu_{b}$. The horizontal black line is located at $\Delta \mu = 0$ in each Galactic coordinate. Purple dots show the mean PM difference in bind of $0.05$ $A_{K_s}^{IRAC}$}.
    \label{fig:ucacgaia_comparison}
\end{figure}

In order to confirm the hypothesis that the problem is in Gaia and not in VVVX, we looker for independent, accurate PMs for this region. To this end, we considered PM values of UCAC5 \citep{Zacharias_2017}. We found $8210$ objects both with PMs both in Gaia DR3 and in UCAC5. Their PM difference was compared to the near-IR reddening $A_{K_s}$ that we obtained through the Rayleigh-Jeans color excess (RJCE) method \citep{Majewski_2011}. Following \citet{Zasowski_2013}, we used the $[4.5 \mu m]$ IRAC band, and the magnitude in the $H$-band, to obtain $A_{K_s}^{IRAC}$ for each star. This comparison is presented in Fig.~\ref{fig:ucacgaia_comparison}. Here, purple dots represent the mean PM difference in bins with $A_{K_s}^{IRAC}$ width of $0.05$ mag. A linear regression was applied to these dots just to highlight that a trend is observed, in the sense of a larger PM difference at larger extinction.\\

We currently do not have an independent way to verify if the correct data are those of the VVVX or those of Gaia. Although the latter represents the state of the art for precision PM, we have presented several independent indications that its PM might be biased at high extinction. We believe that this specific point should be investigated further. Meanwhile, as we could not find reasons to discard our measurements in favor of Gaia, we present possible interpretations of them in the next section.

\subsection{Possible origins of the observed skewness in longitude PM}

\begin{figure}
    \centering
    \includegraphics[width=\columnwidth]{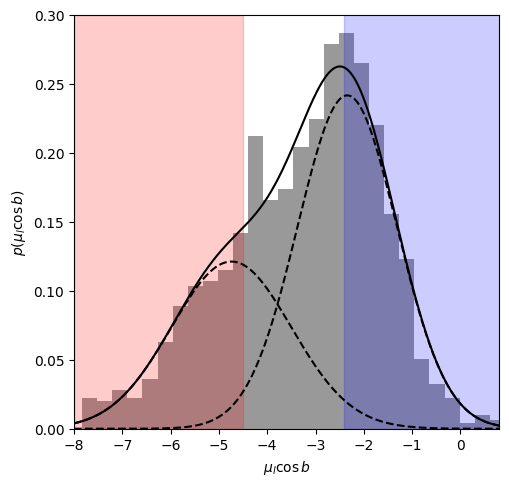}
    \caption{Probability density function of $\mu_{l} \cos b$ with its GMM fit. The two components of the GMM model are displayed (black dashed lines) along with the cuts and "uncertain zones" (not shaded region) adopted to separate the two populations minimizing cross contamination.}
    \label{fig:mul_gmm}
\end{figure}

Assuming that the PM distribution measured from VVVX is correct, we investigate here possible causes of it. The distribution of $\mu_{l} \cos b$ values for stars in our catalogue is shown in  Fig.~\ref{fig:mul_gmm}, where the probability density function of this distribution is depicted. After removing $3\sigma$ outliers from a sigma clipping procedure, we performed a Gaussian mixture model (GMM) fit to the remaining data, which favoured the use of two Gaussian components for this fit (in black dashed lines). These could be associated to two different kinematic populations of stars that cluster around two mean PM values.\\

The observed PM dispersion can be interpreted in two different ways. The first one is related to the age of the stars in our sample. Two populations of stars of different ages could be coexisting in this region due to the triggered star formation process expected in the region. So that, the presence of two different populations in not necessarily odd.\\

To test the age difference hypothesis and in order to look for the origin of the putative two kinematic populations, we imposed limits to define them. First, $550$ stars presented values of $\mu_{l} \cos b \geq -2.4$ mas $yr^{-1}$ and were defined as part of the \textit{blue overdensity} (blue shaded region on Fig.~\ref{fig:mul_gmm}). These are associated with the Gaussian on the right-hand side of the plot. On the other hand, $382$ objects that had $\mu_{l} \cos b \leq -4.5$ mas $yr^{-1}$ where pointed as part of what we defined as the \textit{red overdensity} (red shaded region on Fig.~\ref{fig:mul_gmm}). These are associated with the Gaussian on the left-hand side of the plot.\\

The limits described above were imposed to minimize cross-contamination between the defined overdensities. However, for the limit imposed for the red one, this is very small. Considering only the region where $\mu_{l} \cos b \leq -4.5$ mas $yr^{-1}$ (red shaded in Fig.~\ref{fig:mul_gmm}), $56 \%$ of stars from the left-hand Gaussian will be included. In this same red region, only $2 \%$ of stars from the right-hand Gaussian  will be included. This indicates that our defined red overdensity is a nearly pure population.\\


    \label{fig:mul_distribution}

\begin{figure}
    \includegraphics[width=\columnwidth]{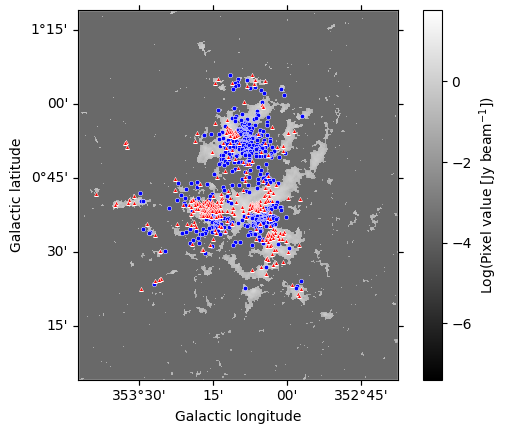}
    \caption{Spatial distribution of all literature sources in VVVX data coloured according to the overdensity they belong to (blue points or in red triangles, according to the limits imposed). These stars are superimposed on the $870$ $\mu$m image of ATLASGAL, indicating the dust continuum emission.}
    \label{fig:spatial_distribution_over}
\end{figure}

As previously mentioned, three main stellar clusters have been identified within NGC 6357, namely Pismis 24, AH03J1725–34.4 and "B cluster" \citep{Massi2015}. Using the limits mentioned above, we can confirm that the different kinematic populations are not correlated preferentially with the projected spatial location of any of the three clusters. The spatial distribution of the two kinematically different populations is shown in Fig.~\ref{fig:spatial_distribution_over}. Here, blue overdensity stars are represented as blue points, whereas the red overdensity is in red triangles and they are plotted over the $870$ $\mu$m image from ATLASGAL \citep{Schuller_2009}. It is observed that objects from both overdensities are present in all the three main stellar clusters of NGC 6357. However, stars from the red overdensity seem to be located in regions where emission from the dust continuum is higher, whereas stars from the blue overdensity are mostly related to areas devoid of dust.\\

\begin{figure}
    \includegraphics[width=\columnwidth]{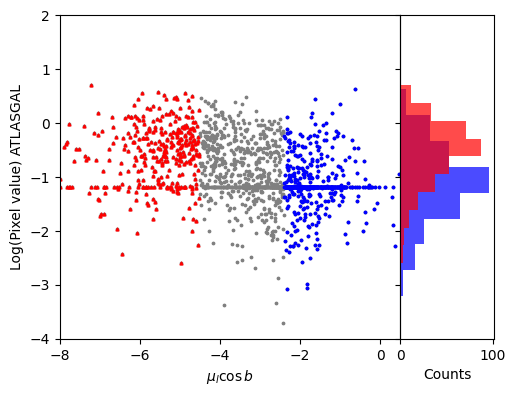}
    \caption{\textit{Left panel}: Distribution of the logarithm of the pixel values from ATLASGAL image in the position of each star. \textit{Right panel}: Histogram of the distribution of the logarithm of the pixel values from ATLASGAL for both overdensities (in blue and red). Bad pixel values are not included.}
    \label{fig:mul_pixel}
\end{figure}

This last is demonstrated and highlighted in Fig.~\ref{fig:mul_pixel}, where the the logarithm of the ATLASGAL flux in the central pixel of each star is shown against $\mu_{l} \cos b$. The horizontal stack of points is related to stars that were located in bad pixels of the image, in which we fixed them to a minimum value for the emission, related to noise. The ones from the red overdensity display higher pixel values than the ones from the blue overdensity, indicating that they are more related to dusty regions. However, as contamination from one overdensity to the other may still be present, a part of these objects are in regions where dust emission is low, as one can observe in the right panel from Fig.~\ref{fig:mul_pixel}, which shows the distribution of the logarithm of the pixel value for each overdensity.\\

\begin{figure*}
    \includegraphics[scale=0.65]{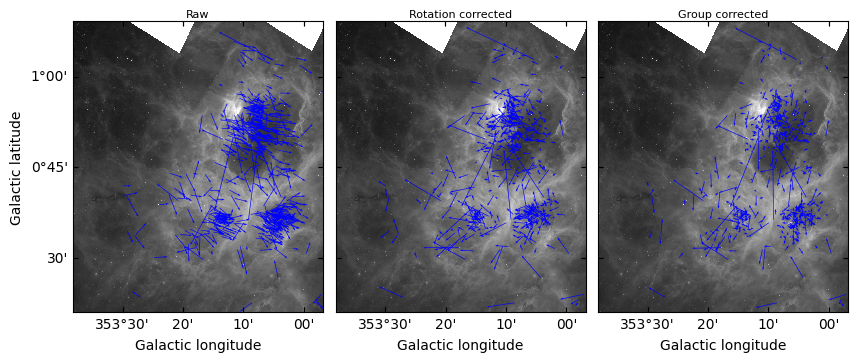}
    \caption{Stars belonging to the blue overdensity superimposed on the Spitzer $8$ $\mu m$ image of NGC 6357 region. Each arrow represents the apparent movement of a given star, according to its values of PMs. \textit{Left panel:} Raw PM values; \textit{middle panel:} PMs without the rotation of the Galaxy; \textit{right panel:} PMs without the mean of the group PM for each coordinate.}
    \label{fig:pms_over1}
\end{figure*}

\begin{figure*}
    \includegraphics[scale=0.65]{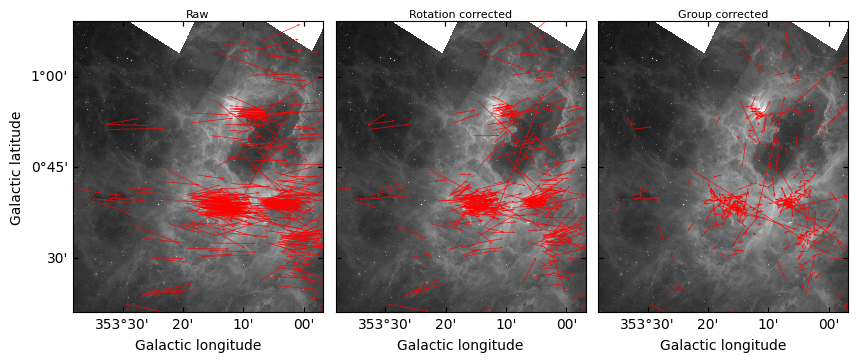}
    \caption{Same as Fig.~\ref{fig:pms_over1}, but now for objects belonging to the red overdensity.}
    \label{fig:pms_over2}
\end{figure*}

To further study the relation between the projected movement of stars and dust regions and to complement the ATLASGAL image, we plotted the PM vectors of stars in each overdensity onto the \textit{Spitzer Space Telescope} GLIMPSE $8$ $\mu m$ mosaic image of higher spatial resolution \citep{Churchwell_2009}. In the latter, filamentary regions are observed and associated to the presence of dust. Raw PMs for the blue overdensity are shown in the left panel of Fig.~\ref{fig:pms_over1}, whereas the ones for the red overdensity are in Fig.~\ref{fig:pms_over2}. Here, the vector lengths are merely orientative, but they all share the same scale. Note that the PM vectors here include the disk rotation and the movement of the Sun around the Centre of the Galaxy.\\

In order to remove the rotation of the Galactic disk, we considered stars within a control region located outside NGC 6357 but at approximately the same Galactic latitude. We built the color-magnitude diagram (CMD) of those objects and selected only the stars that appear to belong to the disk due to their position on this diagram. All of them had magnitudes in the range $10.5 \leq K_s \leq 13$. From them, we considered only the ones that shown PM errors less than the mean of the VVVX PM uncertainties, to assure a good measurement of these quantities. For those objects, that we assume to belong to the disk, we computed their mean PM in each Galactic coordinate ($\mu_{l}\cos b, \mu_{b})= (-2.549, -0.779)$ mas $yr^{-1}$. This was then subtracted from the raw PMs, resulting in what is shown in the middle panel of Fig.~\ref{fig:pms_over1} and Fig.~\ref{fig:pms_over2}. For the blue overdensity, one can observe that their orderly movement is lost, while for the red ovedensity this does not happen. From this, we can infer that stars from the blue overdensity have had more time to interact dynamically and reach a relaxed state. Objects from the red overdensity, even if we subtract the rotation of the Galaxy, keep their ordered motion and they follow the filaments of the region. This suggests that they are a younger population, whose movement is dominated by that of the molecular cloud it is attached to.\\

Furthermore, to investigate the local kinematic substructure within a group (expansion, for instance), we computed the mean of the motion in each coordinate for the two overdensities and subtracted them from the raw PMs. This is shown in the right panel of Fig.~\ref{fig:pms_over1} and Fig.~\ref{fig:pms_over2}. In both cases, the ordered motion of stars is lost. For stars in the red overdensity, they could move as a group because they are in a more dense environment or because they are younger than objects from the blue overdensity.\\

\begin{figure*}
    \includegraphics[scale=0.6]{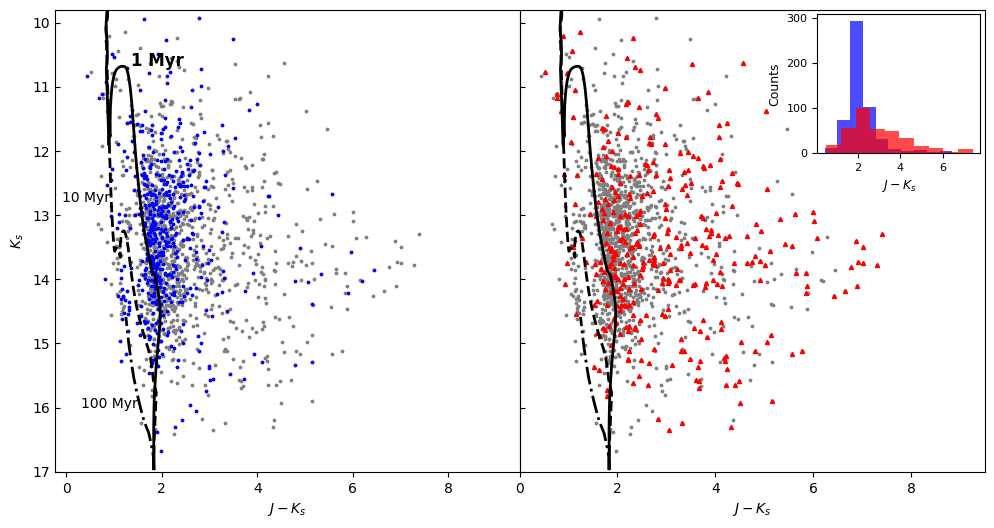}
    \caption{CMDs for the two kinematically different populations of stars (blue overdensity on the left panel and the red one on the right panel). Gray points represent all the stars with available PMs. Pre-Main Sequence Padova isochrones of $100$ (dot dashed line), $10$ (dashed line) and $1$ Myr (thick solid line) from \citet{Bressan_2012} are also shown. These have been reddened considering $Av=5.93$ \citep{Russeil2012}, using the extinction law of \citet{RiekeLebofsky_1985} and located at a distance of $d=1.7$ kpc. The inset panel shows a histogram for the color $J-K_s$ of each overdensity.}
    \label{fig:cmd_over}
\end{figure*}

In order to further characterize similarities and differences of the two possible kinematic populations, we produced a CMD for stars of our catalogue that belong to each kinematic population shown in Fig.~\ref{fig:cmd_over}. Pre-Main Sequence Padova isochrones of $100$, $10$ and $1$ Myr (left to right) from \citet{Bressan_2012} are shown. The latter is highlighted because it represents the age of the region. They have been reddened using $A_V = 5.93$ \citep{Russeil2012} and located at a distance of $1.7$ kpc. Both populations could be associated to young stars in the region. However, sources from the red overdensity seem to have higher color dispersion than the ones from the blue overdensity. This is emphasized in the inset panel, where a histogram for the color $J-K_s$ of each overdensity is presented.\\

As observed in the CMDs in Fig.~\ref{fig:cmd_over}, there is a higher spread in color for red overdensity stars. While members of the blue population are seen closer to the $1$ Myr isochrone, the ones from the red population are more spread along the horizontal axis (also shown in the inset panel of Fig.~\ref{fig:cmd_over}). This can be attributed to two main factors. The first one is the degree of differential extinction of the region, which would make them look redder than they are. The second one is the age of the sources themselves. If these objects are associated to regions in which extinction should be more homogeneous, as filamentary regions could be, then we could relate this spread to their age. As they are further away from the isochrones shown, this could indicate a younger age compared to the ones from the blue overdensity. Nevertheless, the observed color can also be affected by the presence of disks or envelopes surrounding these stars.\\

\begin{figure}
    \includegraphics[width=0.9\columnwidth]{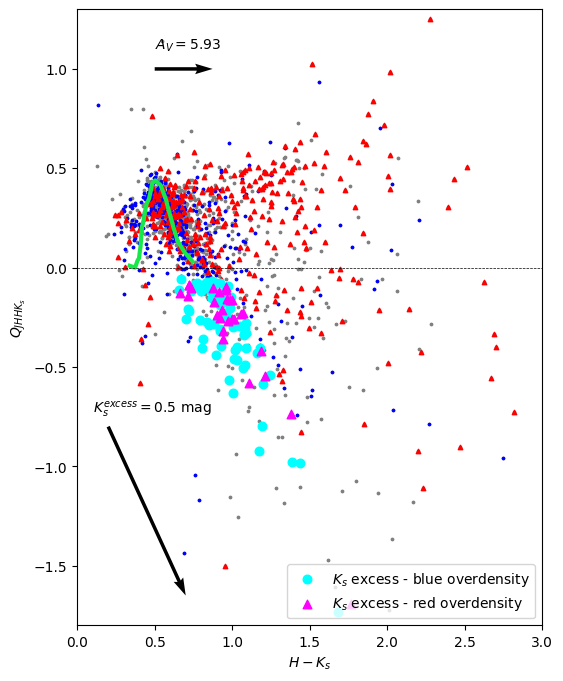}
    \caption{Reddening free index $Q_{JHHK_s}$ vs. $H-K_s$ of stars with available VVVX PMs and colored by the kinematic overdensity they are part of. Stars with $K_s$ excess that are $3\sigma$ away from the mean of the photospheric colors are marked as full cyan circles (magenta triangles) if they are part also from the blue (red) overdensity. The solid green line depicts the location of dwarf stars.}
    \label{fig:Qindex}
\end{figure}

Therefore, to disentangle reddening from the presence of these structures, we have computed the reddening free index $Q_{JHHK_s}$, defined in \citet{Damiani_2006} as:
\begin{equation}
    Q_{JHHK_s} = (J-H) - \frac{E(J-H)}{E(H-K_{s})} (H-K_{s})
\end{equation}
Its values against the $H-K_s$ color, for stars belonging to the two kinematic overdensities, are shown on Fig.~\ref{fig:Qindex}. The location of dwarf stars, only with photospheric colors, is depicted as a green line \citep{Bessel_1988} and its minimum value is represented by the horizontal dashed line. This dwarf location has been reddened considering $A_V=5.93$ mag from \citet{Russeil2012}.\\

The $Q_{JHHK_s}$ index helps to discriminate between stars affected by extinction (located to the right of the dwarf location, along the extinction vector $A_V$) and those with an excess in the $K_s$-band due to the presence of envelopes or disks (located below the dashed horizontal line and along the $K_{s}^{excess}$ vector).\\

First, it is observed here that red overdensity stars are indeed more extincted. On the other hand, following \citet{Damiani_2006}, 
we selected stars $3\sigma$ away from normal photosphere colors and along the $K_{s}^{excess}$ vector, that is, candidates for having envelopes or disks. We obtained that the fraction of these candidates is higher for the blue overdensity (cyan full circles) than for the red one (magenta full triangles), when considering $A_{V}= 5.93$. This finding can argue against the red overdensity being younger that the blue one. However, when we move to higher extinction values, as the red overdensity is, we observed that the fraction of stars with $K_s$ excess from this kinematic population increases. This is further explained on Appendix \ref{sec:Qindex_models}.\\

Again, an age gradient is not completely unexpected, since it could be attributed to a triggered star formation process, which was already been invoked in several papers (see e. g. \citeauthor{Wang2007} \citeyear{Wang2007}, \citeauthor{Cappa2011} \citeyear{Cappa2011}, \citeauthor{Fang2012} \citeyear{Fang2012}, \citeauthor{Massi2015} \citeyear{Massi2015}). However, the origin of this second star formation event is still debated. A big shell in H$\alpha$ data, as well in the DSS R-band, is observed towards NGC 6357 (Fig.~\ref{fig:NGC6357_DSSR}) and it has been associated with a supernova event. Either this or the expansion of the bubble CS 61 have been invoked as the possible origins of the compression of the remaining gas in the region. However, through our data, the expansion of the shell is favoured. The big shell encompasses a larger region than the three clusters of NGC 6357. Therefore, its expansion could push the material in a coherent direction and generate the orderly motion observed in the PMs of the younger stars.\\

Nevertheless, the second plausible explanation for the observed PM behaviour could be related to an asymmetric expansion of the stars. The latter was detected in the Vela OB2 association by \citet{Armstrong_2020}, where two kinematic populations were found. The expansion was confirmed by considering positions, parallaxes and RVs of the stars, in addition to their PMs, to obtain their positions X, Y, Z and velocities U, V, W in the Galactic Cartesian coordinate system. Unfortunately, for great majority of the stars in our sample, RVs are not available in the literature, which does not allow us to confirm whether these stars are also expanding. However, we cannot rule out that this is could be the origin of the two observed kinematic populations, particularly due to the higher dispersion found for the $\mu_{l} \cos b$ PM values.\\

This latter hypothesis is favoured through the fraction of stars that are candidates of having disks or envelopes, found through the $Q_{JHHK_s}$ index. Stars from the blue overdensity have a higher fraction of these candidates and this could rule out an age difference between both kinematic overdensities. Stars with disks are expected to be younger than those without them, which may occur if their evolution is under isolated conditions. Nevertheless, it is important to keep in mind that this region contains some of the most massive stars in the Milky Way. Their presence, in turn, can affect the evolution of circumstellar disks of nearby stars by photoevaporation due to UV radiation of OB stars \citep{Ansdell_2017}. Objects with photoevaporated disks can be as young as the ones with disks \citep{Clarke_2007, Anderson_2013, Winter_2020}. Therefore, an age difference between both kinematic populations could remain plausible.

\section{Light curve classification}
\label{sec:QM}

\subsection{Period search}

Each object in the LitCatVVVX filtered for spatial contaminants had available $K_s$ light curves. Three main filters were applied to them, in order to obtain a catalogue of reliable light curves. First, we kept the ones that had, at least, $70$ data points and, second, we only considered the ones in which their mean $K_s$ magnitude, $\overline{K_s}$, had a value higher than $10.5$ mag. The latter with the aim of avoid saturated sources. Considering this magnitude limit and the $1$ Myr Padova isochrone from \citet{Bressan_2012}, stars of masses $M\approx7 M_{\odot}$ and above will be left out.\\

Besides, as we are interested in real brightness variations, we included a third condition that left only the light curves with median error $eK_s$ less than $10\%$ of the amplitude of the variation $\Delta K_s = K_{s, max} - K_{s, min}$.\\

For all the light curves that met the three conditions mentioned above, we computed the Lomb-Scargle (LS) periodogram of its $K_s$ light curve \citep{Lomb1976, Scargle1982}. As explained in \citet{OrdenesHuanca2022}, the two most probable periods were computed, using the same frequency range of $7.4 \times 10^{-4}$ $d^{-1}$ and $2.5$ $d^{-1}$. Due to the $1^{d}$ cadence of VVVX survey, periods around that value, $0.5$ day and $2$ days could be artificial or aliases. We therefore removed additional $108$ objects having both their most probable periods near these values. It is important to keep in mind here that the ``periods" obtained should be more appropriately called variability timescales, as not all of the light curves show strictly repeating waveforms. The periodicity metric $Q$, as will be explained later in the text, will help us asses the stability of such "periods".\\



Regarding the periods, we treated two cases. First, we considered stars with periods that were less than $40$ days, as it is the typical timescale of variation in young stars. These allowed us to account for $561$ stars that met the criteria mentioned. For stars where the LS periodogram suggested periods longer than $40$ days, we visually inspected their light curves. The ones that showed flux variations at longer timescales were included. For example, some of these stars showed brightness decreases or increases in the entire baseline of observation or presented slower flux variations. These $213$ objects were also included in our list.\\

These criteria left us with a list of $774$ light curves to be classified according to their degree of periodicity $Q$ and asymmetry $M$. The summary of the filters applied and the number of stars left in the list are presented in Table~\ref{tab:filters}.

\begin{table}
	\centering
	\caption{Filters applied to the VVVX light curves with the number of stars that passed each.}
	\label{tab:filters}
	\begin{tabular}{cc} 
		\hline
		Filter & \# stars\\
		\hline
		Young stars with reliable light curves in VVVX & 1786\\
		$70$ data points, no saturation/alias, true variability, $P \leq 40$ days & 561\\
		$P \leq 40$ days + Long timescale & 774\\
		\hline
	\end{tabular}
\end{table}

\subsection{Q and M metrics}

Young stars have light curves associated to flux changes due to different physical mechanisms, as explained in Sect.~\ref{sec:intro}. The stability of the period, as well as its tendency to increase or decrease the brightness will allow us to infer the physical process that dominates the variability observed. Following \citet{Cody_2014}, we considered their two parameters $Q$ and $M$ that serve as a guidance to know this information. These are extensively explained in their work, although for space-based telescopes. Their adaptation to ground-based data, as is our case, is described in detail in the work of \citet{Hillenbrand_2022}. In this work we follow the same procedure and as we did in \citet{OrdenesHuanca2022}.\\


The periodicity parameter $Q$ carries the information regarding the stability of the period and shape of the light curve. In this work we only considered two $Q$ classes. \textit{Quasi-periodic} light curves will be the ones that have a shape that is evolving in time, but the period remains fixed, with $Q 
\leq 0.6$. Conversely, if the period resulted in an unstable one, an \textit{aperiodic} classification will arise and $Q > 0.6$. Therefore, the period obtained is not related to a repeated pattern for those light curves, but to a timescale of variation. In eight years of observations, the light curves in the VVVX catalogue always show evolving patterns, so that, we did not consider the \textit{periodic} class.\\


The "dominant" state, in terms of brightness, of light curves is captured by the $M$ parameter that distinguishes between symmetric variations, dominant "low-state" values with sudden increases, or, the opposite, dominant "high state values" with sudden decreases. \textit{Bursting} light curves have a tendency to increase their fluxes with $M<-0.4$, whereas \textit{dipping} ones tend to decrease their brightness and have $M>0.4$. If there is no tendency, a \textit{symmetric} light curve will be placed with $-0.4 \leq M \leq 0.4$.\\

\begin{figure}
    \includegraphics[width=\columnwidth]{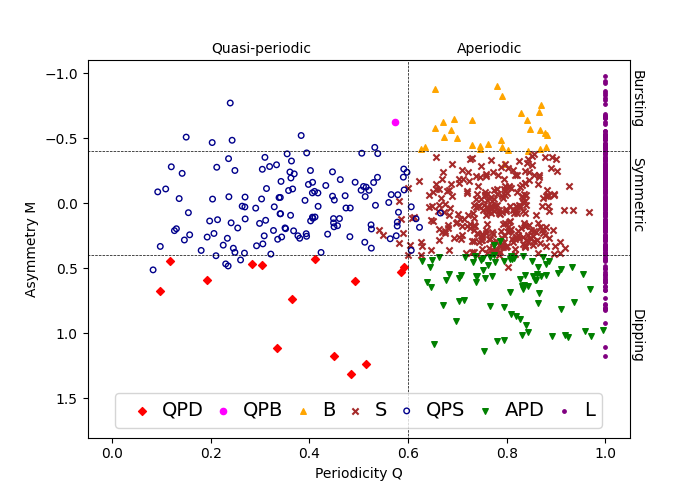}
    \caption{$Q-M$ plane for the $774$ stars in our catalogue. Each color and marker represent the classifications from our visual inspection. (QPS: quasi-periodic symmetric; QPD: quasi-periodic dipping; QPB: quasi-periodic bursting; B: burster; S: stochastic; APD: aperiodic dipping; L: long timescale)}
    \label{fig:QM_all}
\end{figure}

\begin{figure*}
    \centering
    \includegraphics[scale=0.45]{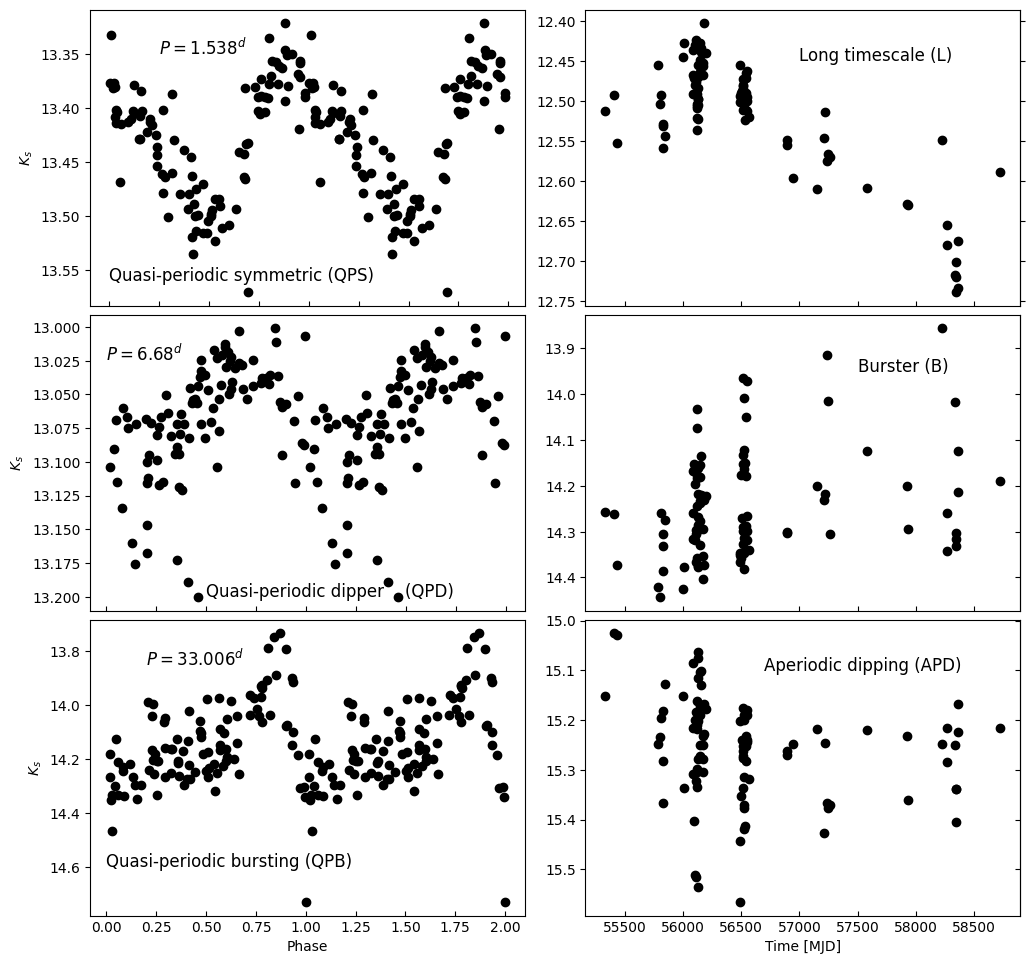}
    \caption{Light curve examples for six of the different classifications according to the metrics $Q$ and $M$. Each class is indicated inside the plots. Only those with stable periods are phase folded (left hand panels) and their values are indicated on each light curve.}
    \label{fig:LC_examples}
\end{figure*}

These two metrics were computed and combined for our $774$ light curves. In addition, as in \citet{OrdenesHuanca2022}, long timescale stars (L) have a fixed periodicity of $Q=1$. The metrics and classes of the light curves in our catalogue are plotted in the $Q-M$ plane, which shows the different categories as separate regions of the plot shown in Fig.~\ref{fig:QM_all}. Examples for the majority of the categories are provided on Fig.~\ref{fig:LC_examples}.\\

It is important to highlight that both metrics are a guidance to infer the process that can explain the largest amount of variability. However, it is also important to visually inspect the light curves to confirm that what is measured for these parameters is actually related to the behaviour of the light curve. Contradictory cases could appear especially at the limits of each classification. In Fig.~\ref{fig:QM_all} are also included our classifications by eye in colors and markers, which are in very good agreement with the metric values for most of stars. Here we introduced the class QPB for one source, which is related to a quasi-periodic bursting variation shown in the bottom left panel of Fig.~\ref{fig:LC_examples}. In this case, pulsed accretion could be the dominant source of flux change.\\

As can be observed, the majority of stars are related to aperiodic behaviours (middle and bottom right panels of Fig.~\ref{fig:LC_examples}). In eight years of observations, variability due to spots are the ones that can maintain a fixed value for the period, as shown in the upper left panel of Fig.~\ref{fig:LC_examples}. This, because it is linked to the magnetically active regions of the star associated to its rotation. Aperiodic behaviors are more common because, as also mentioned in \citet{OrdenesHuanca2022}, when accretion is present, spots can dominate the variability only in seasons or fractions of the entire baseline of observation. Furthermore, if the amplitude due to spots is small enough, it won't be observed as a repeatable pattern, leading to stochastic flux variations.\\

In terms of the asymmetry $M$ metric, the bulk of stars have symmetric light curves. This is attributed to the cadence of VVVX data, which only allow us to observe a lower limit of bursts or dips events. If these occur in less than a day, the probability to observe them is lower. Still, some of these are observed in the VVVX data, but they need to have the sufficient amplitude to be detected in the near-IR. In addition, and as mentioned above, it is expected that a part of the classifications change in time \citep{McGinnis_2015, Ansdell_2020}. For example, a light curve that presents bursts in a fraction of the observation can start to have dips in the next season of observation. This means that, overall, no tendency either to increase or decrease the flux will be measured. In eight years of observation, this can also be more frequent.\\

The near-IR parameters of stars in our light curve catalogue along with their light curve metrics and classes are listed on Table \ref{tab:lcs} from Appendix \ref{sec:appendix1}.\\

\section{Discussion}
\label{sec:discussion}

YSOs variability classes have been related to the age of the stars \citep{Cody_2022}. For instance, QPD stars are expected to become more numerous as the cluster age increases, whereas APD variations should decrease. As a star ages, its accretion rate will decrease, which favours a stable accretion mode \citep{McGinnis_2015}. This can lead to a QPD classification. In addition, accretion processes should be dominant in the flux variability of younger stars with quasi-periodic bursts more related to this population and due to their higher accretion rates. Besides, the amplitude of variability should decrease for older clusters. Although what we observe in the light curves, and subsequently classify, could be affected by the inclination of the stars, they can still provide indications of any age difference in the stellar content of a given region.\\

YSOs can be related to objects with envelopes, disks, or even without any surrounding structure. The presence, or not, of such structures is also linked to the age of these objects. A protostar arises embedded in an envelope and, as it evolves, it develops an accretion disk. Accretion processes and outflows dissipate the envelope and, later, the disk, leading to a pre-Main Sequence star with a thin or no surrounding structure. When an envelope and/or disk is present, it will cause an IR excess that changes the SED morphology.\\

Based on the latter, \citet{Lada_1987} proposed a classification scheme in which the IR excess is related to the slope of the SED in the IR region. This designation was later complemented with the one of \citet{Andre_1993}, who found younger, sub-mm objects embedded in cold dust. Depending on the value of this slope, different YSO classes are defined. Objects that are still embedded in an infalling envelope are the ones that compose \textit{class 0}, whereas \textit{class I} stars are surrounded by an envelope and a circumstellar disk. These are not detected in optical bands, linking them to protostars. On the other hand, \textit{class II} objects are the ones with a surrounding disk and they have been associated with CTTSs. Finally, sources categorized as \textit{class III} have very low-mass or no disks. As a young star evolves, it will pass from being a class 0 source to a class III. Therefore, the fraction of stars in each class can roughly be linked to the age of a given region.\\

YSOs classes have also been analyzed according to their X-ray luminosity. Coronal emission, magnetic flares, as well as strong stellar winds are responsible for the young stars emission at this wavelength. Coronal emission has been related to the so-called soft X-ray spectra, which is in the range of $\sim 0.5 - 2$ keV. A very soft emission $<0.3$ keV has been attributed to accretion. On the other hand, magnetic flares, giving rise to reconnection events in magnetic loops, have been related to the hard X-ray emission, defined in the $2 - 8$ keV energy range \citep{Getman_2005A}. In \citet{Prisinzano2008}, the authors define two YSO subclasses to study their X-ray emission and, subsequently, compare it with that of YSOs of other classes. Class 0/Ia is related to stars with SEDs that monotonically increased from the K-band to $8$ $\mu$m. Objects defined as being part of class 0/Ib presented increasing SEDs from the K-band to $4.5$ $\mu$m and decreasing at longer wavelengths. In this work, the authors claimed that stars in younger stages of evolution, such as class 0/I sources, are less luminous in X-rays than class II/III objects, both in total and hard regimes. However, this was particularly true for objects designated as class 0/Ia. Class 0/Ib stars shown similar X-ray luminosities than class II sources. Further, accreting TTSs have shown to be, on average, less luminous in X-rays than "nonaccreting", class III ones \citep{Preibisch_2005}.\\

In this section, we link the light curve categories, YSO classes and their X-ray luminosities to the two kinematic population of stars found in this work, to check if this age difference is also observed through these parameters.\\

\subsection{Overdensities and light curve classification}
\label{sec:over_class}

\begin{figure*}
    \includegraphics[scale=0.65]{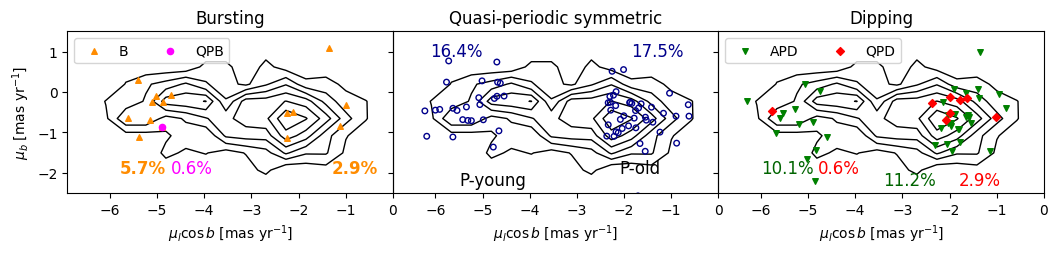}
    \caption{Vector point diagram of stars in our light curve catalogue. Contours highlight the two kinematic populations and each is labeled on the middle panel. Colors indicate their $Q$ and $M$ classifications, as Fig.~\ref{fig:QM_all}: B, QPB (left panel), QPS (middle panel), APD and QPD (right panel). The fraction of each class in the different populations is also denoted and colored accordingly.}
    \label{fig:mul_qm}
\end{figure*}

Using VVVX data we are able to obtain only lower limits for burst or dip events, as mentioned. However, the two overdensities defined in Sec.~\ref{sec:pms} are affected by the same limitation, thus enabling a relative comparison. Considering the same limits imposed above for each overdensity, we found that $240$ objects of our light curve catalogue are part of the blue population, whereas $158$ are of the red population. The VPD for them is shown in Fig.~\ref{fig:mul_qm} and contours highlight the presence and location of the two kinematic populations in the light curve sample. Using them, we looked for a difference in the number of stars that have, first, a bursting behaviour (left panel of Fig.~\ref{fig:mul_qm}) and try to relate them to a difference in age. For the putative old (\textit{p-old}) population, the blue one, $2.9 \pm 1.1 \%$ of the objects are part of this class, while $5.7 \pm 2.0\%$ of stars in the putative young (\textit{p-young}) or red overdensity are bursters. Furthermore, the only QPB star in our sample is part of this likely younger population. In terms of the amplitude of their brightness changes, we could not find a difference in the mean $\langle \Delta K_{s}\rangle$ of burster stars in each overdensity.\\

Stars classified as QPS are expected to have flux changes due to the presence of spots in their surface. This dominates the variability mostly in Class III objects or weak-line T Tauri stars (WTTSs) with weak or no disks. These are expected to be more evolved than disked stars, in which usually accretion is the main source of variability. In our case, we found that both overdensities have a similar number of QPS objects, with $17.5 \pm 2.9 \%$ in the p-old population and $16.4 \pm 3.5 \%$ for the p-young one (middle panel of Fig.~\ref{fig:mul_qm}). Therefore, we cannot draw a difference in age linked to the fraction of QPS stars. However, this region has several massive stars than can affect the evolution of the disks due to photoevaporation.\\

For stars categorized as APD, we found that both overdensities have a similar fraction of objects being part of this class. Around $11.2 \pm 2.3 \%$ for the p-old population, whereas the p-young has $10.1 \pm 2.6 \%$. However, QPD stars are mostly related to the p-old population, with $2.9 \pm 1.1\%$, whereas the p-young overdensity has only one QPD star which represents the $0.7 \pm 0.6 \%$. It is important to mention that all these ratio values could be biased by completeness. The vector point diagram for both classes related to dipping behaviors are shown in the right panel of Fig.~\ref{fig:mul_qm}.\\

The results mentioned above, particularly for the number of stars classified as B, QPB and QPD, point to a possible difference in age and observed through the $Q$ and $M$ metrics. The p-young population have a higher fraction of bursters (B) and the only QPB in our sample, whereas the p-old population has more QPD stars, expected for older accreting objects. However, they are not statistically significant, so we cannot conclusively say that there is an age difference between the two populations. In addition, this has to be taken carefully, because, on one hand, contamination from one population to the other may be present. On the other hand, the dips observed through VVVX data could also be associated to binaries, about which we do not have information. Still, due to one of our filters to compose the catalogue, we know that, at least, these stars do not have close companions. Therefore, the $Q$ and $M$ metrics cannot confirm the age difference as the origin of both populations, but they cannot rule it out either. The latter, particularly because the number of stars in each class is small, affecting the statistics. The origin of both populations due to expansion is still likely.

\subsection{YSO classes and overdensities}
\label{sec:over_ysos}

\begin{figure*}
    \centering
    \includegraphics[scale=0.55]{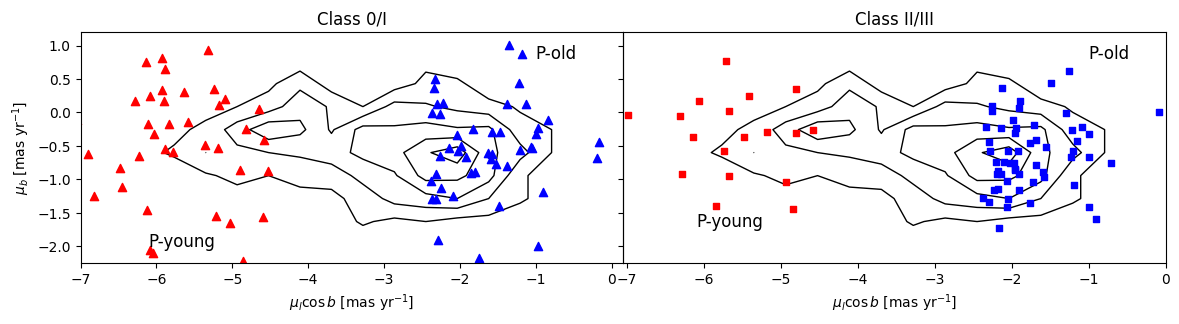}
    \caption{Vector point diagram of stars in our PM catalogue. Contours highlight the two kinematic populations and both are labeled on the plots. Stars with available YSO classification are coloured depending on the overdensity they belong to. \textit{Left panel:} class 0/I stars in red/blue triangles. \textit{Right panel:} class II/III stars in red/blue squares.}
    \label{fig:pms_0II}
\end{figure*}

The MIRES catalogue \citep{Povich2013} is based on the analysis of the SEDs and the presence of an IR excess. Here, the authors give rough classes for the YSOs in their sample, using photometry from $1 - 8$ $\mu$m and the extrapolation redward of the SED. Objects identified as likely YSOs have been classified either as class 0/I or class II/III, all of them showing an IR excess. We considered these classes to look for rough difference between the fraction of sources in each class in a given kinematic population. This is presented in Fig.~\ref{fig:pms_0II}. Again, contours emphasize the two kinematic populations, whereas stars with available YSO class are coloured according to the overdensity they belong. In the left panel of the plot, only class 0/I objects are coloured. In the p-young one, class 0/I sources constitute $17.7 \pm 2.4$ \%, whereas $10.2 \pm 1.4$ \% of the p-old population objects are part of this class.\\

In the right panel of Fig.~\ref{fig:pms_0II}, class II/III stars are shown in coloured squares. Only $\sim 8.8 \pm 1.6$ \% of this type of objects are part of the p-young population, whereas a higher fraction of class II/III belongs to the p-old population, which is $\sim 12.5 \pm 1.6$ \%. Again, as mentioned in the previous subsection, these values have a difference that is not statistically significant and may be affected by the completeness of the data. Still, the values themselves point to a difference in age between both populations. Due to the statistical significance, we cannot yet rule out either of the two possible origins of the two kinematic populations.\\

\subsection{Kinematic populations and their X-ray luminosity}
\label{sec:over_xrays}

\begin{figure}
    \includegraphics[width=\columnwidth]{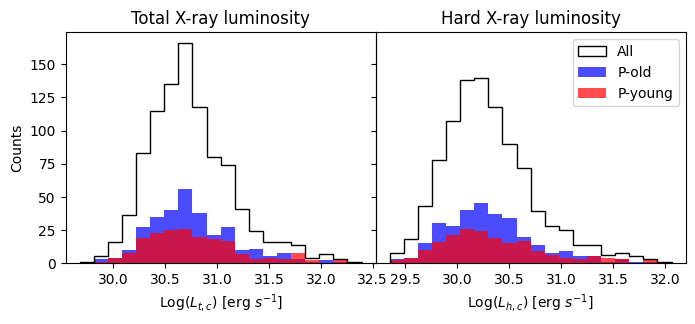}
    \caption{\textit{Left panel}: Distribution of the logarithm absorption corrected total X-ray luminosity Log$(L_{t,c})$ of stars in our PM catalogue (black), coloured by the overdensity they belong. \textit{Right panel}: Same as left panel, but now for the absorption corrected hard X-ray luminosity Log$(L_{h,c})$.}
    \label{fig:lx}
\end{figure}

In the catalogue of MYStIX, the authors give the total ($0.5 - 8$ keV) and hard ($2-8$ keV) X-ray luminosity for the stars in their sample, including the absorption correction, $L_{t,c}$ and $L_{h,c}$, respectively. From our catalogue of PMs, $955$ stars had X-ray luminosities measured. Their distribution of both X-ray emission regimes is shown as the black line histogram of Fig.~\ref{fig:lx}. In this plot, the distribution of $L_{t,c}$ is presented in the left panel, whereas the right panel shows $L_{h,c}$. In both, the kinematic populations are also depicted. For the p-old one, $304$ objects had X-ray luminosity values available, while, $190$ from the p-young population had these data.\\

From the total X-ray luminosity, we can observe a flat distribution for the p-young population (red), whereas the p-old one (blue) presents a peak around Log$(L_{t,c})=30.7$. This peak could be associated with more evolved class III stars. However, the total X-ray luminosity can be affected by extinction. The soft component of the coronal emission ($\sim 0.8-1$ keV) could not be detected in heavily extincted sources, as class 0/I objects are, so their $L_{t,c}$ could be underestimated \citep{Prisinzano2008}. Therefore, including the hard X-ray luminosity is crucial. In our case, from the right panel of Fig.~\ref{fig:lx}, we can observe that there is no obvious difference in Log$(L_{h,c})$ between both kinematic populations. We note here that, despite the fact that the red population has a higher fraction of class 0/I stars, these are not further divided into 0/Ia and 0/Ib classes. The latter was found to have similar X-ray luminosity than class II objects.






\section{Conclusions}
\label{sec:conclusions}

An age difference was expected in NGC 6357. First, multiple or sequential star formation events can take place in open clusters \citep{Massi2015}. Still, if this gradient in age is less than $1$ Myr, it would be a real challenge to determine if it is an actual behaviour or it could only be attributed to errors in age measurements. The current evolutionary tracks for young stars are not reliable enough to confirm an age dispersion at this timescale. Furthermore, several indicators of triggered star formation have been found in this site, such as shocks and gas clumps \citep{Westmoquette2010, Russeil2010}. However, the three clusters that compose NGC 6357, have proven to be nearly coeval, with ages between $1 - 1.5$ Myr \citep{Getman_2014}.\\

A greater age difference was noted in the work by \citet{Russeil_2017}. Here, the authors found that two star-formation events could have taken place on NGC 6357. Using OB stars from this site, they found that $2$ of these massive stars have ages of $\sim 1.4$ Myr, whereas $13$ objects are $\sim 4.6$ Myr old. Therefore, they conclude that a first star-formation event happened $4.6$ Myr ago. Then, a second and main burst occurred at approximately $1.4$ Myr ago, in agreement with was was found on the work by \citet{Getman_2014}.\\

The particular origin of the star-formation event related to stars in our sample has been discussed further in the literature. As shown in Fig.~\ref{fig:NGC6357_DSSR}, a big optical shell is present in a large-scale image of NGC 6357 (also called as the \textit{ringlike nebula} in \citeauthor{Wang2007} \citeyear{Wang2007}, the \textit{big shell} in \citeauthor{Cappa2011} \citeyear{Cappa2011} or the \textit{ring} in \citeauthor{Massi2015} \citeyear{Massi2015}). This shell encompasses a cavity, which has also been labeled as the bubble CS 61 (or G353.12+0.86) and related to Pismis 24.\\

In the work by \citet{Wang2007}, the authors discuss that CS 61 could have been originated through a supernova event. This mainly because of two reasons. The first one is related to the the fact that Pismis 24 is located at the north, and not at the center, of the bubble. So that, its OB stars could not produce such a structure. Second, because of the nearby presence of a Wolf-Rayet star (namely WR 93), they deduce that a sufficiently older population of stars in NGC 6357 was capable of producing such an energetic event that originated this bubble.\\

However, other authors have proposed that this bubble was created through a \textit{champagne flow}, in which the molecular material was pushed to the south due to the strong UV field of the massive stars in Pismis 24 \citep{Giannetti2012}. In the work by \citet{Russeil_2017}, the probable first star-formation event in the NGC 6357 region has also been pointed as a possible creator of the bubble and the filamentary regions in its vicinity. \\



The presence of the bubble is important for triggered star-formation processes. Massive stars from Pismis 24 and their strong winds can lead to its expansion and subsequent compression of the nearby molecular material, which surrounds the optical shell. In \citet{Russeil2010}, a number of dense cores were observed towards the molecular region, suggesting an ongoing star-formation process. \\ 

We have studied the stellar components of NGC 6357 both kinematically and through their flux variability in the $K_s$-band, using data from the VVVX survey. We have found that NGC 6357 is composed of two kinematically different populations of stars. These two have not been identified separately in the literature mainly because of the high extinction of the region which limits optical studies.\\

As observed through the ATLASGAL map, one of the populations, namely the red overdensity, is spatially related to filamentary regions. The projected movements of these objects seem to follow the filaments, as shown in Fig.~\ref{fig:pms_over2}. These two findings would indicate that it could be a putative younger population (\textit{p-young}) than the one denoted as the blue overdensity (or \textit{p-old}).\\ 

Further, traces of an age difference of the two populations are observed through their $Q$ and $M$ values, for a subset of stars. Particularly, for the fraction of objects classified as B, QPB and QPD in each overdensity. In addition, the fraction of objects classified as class 0/I in the p-young population is higher than the one found for the p-old one, which could also indicate that the red overdensity would be younger. However, the values obtained are not statistically significant, so we cannot conclusively say that they show an actual difference in age. In addition, we could not find any difference between their total and hard X-ray luminosities. If the difference in age hypothesis is true, we cannot obtain how much is it from our data alone. By analyzing the CMDs and the evolutionary tracks, we can observe that stars from the blue overdensity have ages around $1$ Myr, whereas stars from the red overdensity seem to be younger, although the positions of red overdensity stars in the CMD could be highly affected by extinction. Neither of these two populations would be related to the one of $\sim 4.6$ Myr identified in the work by \citet{Russeil_2017}, but our p-old objects can be related to their $1.4$ Myr old stars, which is also consistent with the age range found by \citet{Getman_2014} of $1 - 1.5$ Myr. However, due to the uncertainties quoted in these studies, the methods used shall not be sensitive to age differences of less than $\approx 0.5$ Myr, which may be our case.\\

Furthermore, according to the findings for the $Q_{JHHK_s}$ index, the blue population would not be older, considering that the different classes of YSOs follow an evolutionary scheme, from class 0 to class III. Nevertheless, it is important to keep in mind that the evolution of protoplanetary disks is especially affected by the presence of massive stars in this region, which can photoevaporate these structures. Therefore, we cannot discard an age difference as the origin of both kinematic overdensities.\\

The ongoing star-formation process pointed in \citet{Russeil2010} is in agreement with our identification of a likely younger population of stars spatially coincident with the filamentary area. The origin of the shell and the bubble are still uncertain. However, the coherent motion found for the red overdensity of stars points to the fact that the expansion of the big shell could have led to such an orderly movement. This, in turn, may be related to another possible origin of the two kinematic overdensities found. The asymmetric expansion of the members of this region may also cause a higher dispersion in the $\mu_{l}\cos b$ values than the one of $\mu_{b}$, so we cannot rule out that this may be the source of the observed kinematic behavior and not necessarily an age difference. Further studies need to be done, particularly related to the measurement of RVs and parallaxes which can complement our data and prove whether expansion is the real cause of our results. From our findings, both interpretations are equally possible.\\

The two hypothesis mentioned above are based on the PM asymmetry observed through the VVVX data. However, this was not observed on Gaia VPD. Better data are needed (maybe another couple of epochs in a few more years) to clarify whether the VPD of VVVX is correct or not. Maybe even Gaia DR4 can clarify this discrepancy in the future.\\



\section*{Acknowledgements}

C.O.H. acknowledges the support from National Agency for Research and Development (ANID), Scholarship Program Doctorado Nacional 2018–21180315 and ANID Millennium Institute of Astrophysics (MAS) PhD Scholarship. This project was funded by ANID FONDECYT Regular 1230731, ANID Millennium Institute of Astrophysics (MAS) under grant ICN12\_009, the ANID BASAL Center for Astrophysics and Associated Technologies (CATA) through grants AFB170002, ACE210002 and FB210003, and ANID Millennium Nucleus for Planetary Formation (NPF) (grant NCN19\_171), the Max Planck Society (“Partner Group” grant) and the Deutsche Forschungsgemeinschaft (Germany's Excellence Strategy – EXC 2094 – 39078331).

We thank Loredana Prisinzano for her very useful comments, which helped to strengthen the arguments of this study.

We gratefully acknowledge the use of data from the VVV ESO Public Survey program ID 179.B-2002 taken with the VISTA telescope, and data products from the Cambridge Astronomical Survey Unit (CASU). The VVV Survey data are made public at the ESO Archive. Based on observations taken within the ESO VISTA Public Survey VVV, Program ID 179.B-2002. 

\section*{Data Availability}

The VVV and VVVX data are publicly available at the ESO archive \url{http://archive.eso.org/cms.html}. Light curves and PMs, obtained through PSF photometry of VVVX data, have not yet been publicly released but are available on request to the first author.



\bibliographystyle{mnras}
\bibliography{example} 




\appendix

\section{Parameters of YSOs with available PM values in VVVX data}
\label{sec:appendix0}
Table \ref{tab:pms} lists the parameters for the young stars with available PM values in VVVX data.

\begin{table*}

    \caption{List of YSOs with measured VVVX PMs along with their galactic coordinates; their photometric data, $J$ magnitude, its error $eJ$, $H$ magnitude, its error $eH$, $K_s$ magnitude, its error $eK_s$; their kinematic data, Galactic longitude PM value $\mu_{l} \cos b$, its error $e\mu_{l} \cos b$, the Galactic latitude PM value $\mu_{l}$ and its error $e\mu_{l}$ are the columns of the table.}
    \label{tab:pms}
    \resizebox{\textwidth}{!}{%
    \begin{tabular}{|l|c|c|c|c|c|c|c|c|c|c|c|c|}
\hline
        ID & Galactic longitude [deg] & Galactic latitude [deg] & $J$ & $eJ$ & $H$ & $eH$ & $K_{s}$ & $eK_{s}$ & $\mu_{l} \cos b$ & $e\mu_{l} \cos b$ & $\mu_{b}$ & $e\mu_{b}$ \\ \hline
        b328\_513\_13098 & 352.930216 & 0.609676 & 14.2662 & 0.0032 & 13.4426 & 0.0065 & 13.0889 & 0.0014 & -2.775 & 0.34 & -0.583 & 0.34 \\ \hline
    b328\_513\_15831 & 352.796024 & 0.601228 & 19.4352 & 0.0466 & 15.3406 & 0.0088 & 13.3499 & 0.0014 & -7.351 & 0.32 & -0.403 & 0.25 \\ \hline
    b328\_513\_20687 & 352.805617 & 0.585606 & 19.2672 & 0.0436 & 16.7556 & 0.0213 & 14.5809 & 0.0028 & -4.186 & 0.6 & -1.259 & 0.49 \\ \hline
    b328\_513\_16540 & 352.748078 & 0.599246 & 19.4072 & 0.0464 & 16.6946 & 0.0177 & 15.2189 & 0.0044 & -7.562 & 0.91 & -0.081 & 0.92 \\ \hline
    b328\_513\_22367 & 352.757852 & 0.580692 & 18.6422 & 0.0212 & 15.3276 & 0.0092 & 13.7159 & 0.0016 & -10.308 & 0.36 & -3.619 & 0.29 \\ \hline
    \end{tabular}%
    }
\textbf{Note}. A part of Table \ref{tab:pms} is shown here for guidance regarding its content and structure. The entire table is published in the machine-readable format.

\end{table*}

\section{Near-IR reddening free index}
\begin{figure*}
\label{sec:Qindex_models}
    \includegraphics[scale=0.55]{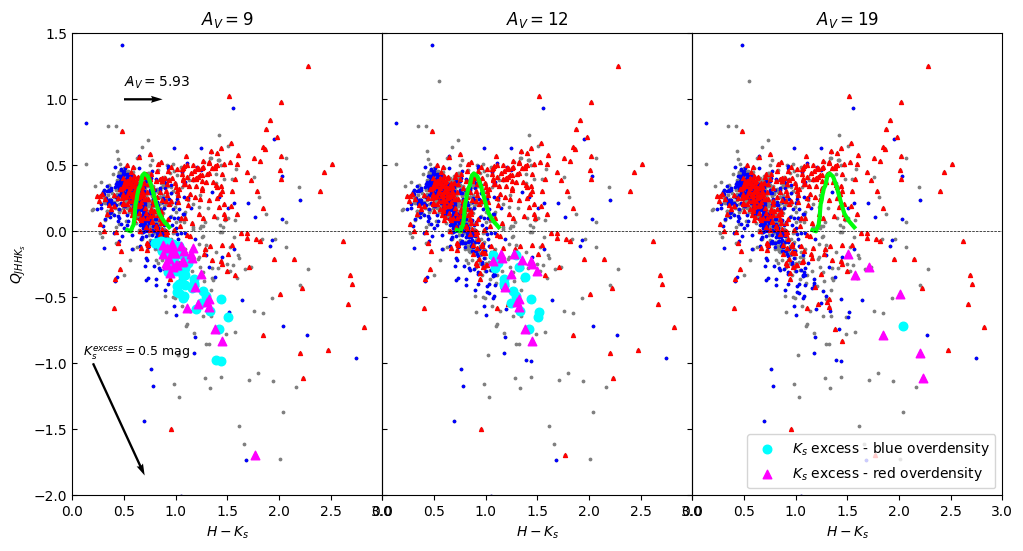}
    \caption{$Q_{JHHK_s}$ vs. $H-K_s$ for different values of $A_V$ considered. The dwarf location according to the extinction value is shown as a green line and its minimum value is marked as a horizontal dashed line. Stars $3\sigma$ away from normal photospheric values of each model are marked in cyan dots or magenta triangles, depending on the kinematic overdensity they belong to.}
    \label{fig:Q_models}
\end{figure*}

In addition to the reddened model with $A_{V}=5.93$, shown in Fig.~\ref{fig:Qindex}, the plots for the reddening free index against the $H-K_s$ values are presented in Fig.~\ref{fig:Q_models}, considering different $A_V$ values that could also fit our data. Particularly, for $A_{V}=9$ (left panel), $12$ (middle panel) and $19$ (right panel). These values are proposed by visually inspecting this plot. The dwarf location, reddened according to the corresponding extinction vector, is also shown in each panel (green line).\\

In order to compute a fraction of star candidates having envelopes or disks, we considered the following. First, we selected stars that were above the horizontal dashed line and between a color bin defined by the minimum and maximum color of the reddened dwarf location. With these, we computed a mean for the $Q_{JHHK_s}$, representative of stars showing only photospheric colors. Then, stars that were $3\sigma$ away from this mean and along the $K_{s}^{excess}$ vector (below the horizontal dashed line), were considered as candidates for having envelopes or disks.\\

We observed that the fraction of stars with $K_s$ excess from the red overdensity increases as they become more extincted. For each $A_V$ value, the fraction of $K_s$ excess objects belonging to each overdensity is presented on Table \ref{tab:Q_fractions}.\\

\begin{table*}
\caption{Fraction of stars, including Poissonian errors, considered as candidates for having an envelope or disk for each dwarf model considered.}
\label{tab:Q_fractions}
\resizebox{0.3\textwidth}{!}{%
\begin{tabular}{|l|c|c|}
\hline
$A_V$ & $K_s$ excess - blue & $K_s$ excess - red \\ \hline
$5.93$ & $0.41 \pm 0.06$ & $0.16 \pm 0.03$ \\ \hline
$9$ & $0.36 \pm 0.06$ & $0.20 \pm 0.04$ \\ \hline
$12$ & $0.31 \pm 0.08$ & $0.27 \pm 0.08$\\ \hline
$19$ & $0.06 \pm 0.06$ & $0.39 \pm 0.17$ \\ \hline
\end{tabular}%
}
\end{table*}

\section{Parameters of YSOs in our light curve catalogue}
\label{sec:appendix1}
Table \ref{tab:lcs} lists the physical parameters for the young stars in our catalogue of light curves.

\begin{table*}

    \caption{List of YSOs within our VVVX light curve catalogue with their galactic coordinates, the mean $K_{s}$ magnitude and its mean error $\overline{eK_{s}}$, the periods obtained, the amplitudes of variation $\Delta K_{s}=K_{s, max}-K_{s, min}$ without outliers, the metrics for the classification of the light curves, $Q$ and $M$, and our classification by visual inspection.}
    \label{tab:lcs}
    \resizebox{\textwidth}{!}{%
    \begin{tabular}{|l|c|c|c|c|c|c|c|c|c|}
    \hline
        ID & Galactic longitude [deg] & Galactic latitude [deg] & $\overline{K_{s}}$ & $\overline{eK_{s}}$ & $P(K_{s})$ [days] & $\Delta K_{s}$ & Q & M & Class \\ \hline
        b329\_201\_718 & 353.106055 & 0.650226 & 14.854 & 0.045 & 0.556 & 0.656 & -0.355 & 1.402 & S \\ 
        b329\_201\_1585 & 353.107178 & 0.647513 & 12.298 & 0.031 & 0.475 & 0.783 & 0.32 & 0.478 & S \\ 
        b329\_201\_1707 & 353.091058 & 0.646978 & 14.025 & 0.032 & 0.299 & 0.651 & -0.086 & 3.867 & S \\ 
        b329\_201\_1866 & 353.110169 & 0.646683 & 13.458 & 0.028 & 0.645 & 0.856 & -0.377 & 0.477 & S \\ 
        b329\_201\_2028 & 353.099518 & 0.646092 & 13.541 & 0.019 & 0.232 & 0.515 & -0.123 & 6.129 & QPS \\ \hline
    \end{tabular}%
    }
\textbf{Note}. A part of Table \ref{tab:lcs} is shown here for guidance regarding its content and structure. The entire table is published in the machine-readable format.

\end{table*}


\bsp	
\label{lastpage}
\end{document}